\documentclass[10pt]{iopart}

\usepackage{iopams}
\usepackage{bm}
\usepackage{iopams}
\usepackage{float}
\usepackage[]{graphicx} 
\usepackage[upright]{fourier}
\usepackage{hyperref}
\usepackage[font=small,labelfont=bf]{caption}
\usepackage[nameinlink,noabbrev]{cleveref}
\usepackage{xcolor,soul}
\hypersetup{
	colorlinks,
	linkcolor={gray!10!blue},
	citecolor={gray!10!blue},
	urlcolor={gray!10!blue}
}

\newcommand*{\figref}[2][]{%
	figure
	\hyperref[{fig:#2}]{%
		\ref*{fig:#2}%
		\ifx\\#1\\%
		\else
		~(#1)%
		\fi
	}%
}

\newcommand*{\Figref}[2][]{%
	Figure
	\hyperref[{fig:#2}]{%
		\ref*{fig:#2}%
		\ifx\\#1\\%
		\else
		~(#1)%
		\fi
	}%
}

\newcommand*{\Eqref}[2][]{%
	equation
	\hyperref[{eq:#2}]{%
		\ref*{eq:#2}%
		\ifx\\#1\\%
		\else
		~(#1)%
		\fi
	}%
}

\newcommand*{\EQREF}[2][]{%
	Equation
	\hyperref[{eq:#2}]{%
		\ref*{eq:#2}%
		\ifx\\#1\\%
		\else
		~(#1)%
		\fi
	}%
}

\newcommand*{\Secref}[2][]{%
	section
	\hyperref[{sec:#2}]{%
		\ref*{sec:#2}%
		\ifx\\#1\\%
		\else
		~(#1)%
		\fi
	}%
}

\begin{document}

\title[Helicon waves in a converging-diverging magnetoplasma]{Helicon waves in a converging-diverging magnetoplasma}

\author{F Filleul$^1$, A Caldarelli$^1$, K Takahashi$^2$, R W Boswell$^3$, C Charles$^3$, J E Cater$^3$, N Rattenbury$^1$}

\address{$^1$Department of Physics, The University of Auckland, Auckland, 1010, New Zealand}
\address{$^2$Department of Electrical Engineering, Tohoku University, Sendai, 980-8579, Japan}
\address{$^3$Space Plasma, Power and Propulsion Laboratory, Research School of Physics, The Australian National University, Canberra, ACT 2601, Australia}
\address{$^4$Department of Mechanical Engineering, University of Canterbury, Christchurch, 8140, New Zealand}
\ead{felicien.filleul@auckland.ac.nz}

\begin{abstract}
Waves propagating along a converging-diverging rf magnetoplasma having the characteristics of a bounded $m=0$ helicon mode are reported and characterised. The discharge features a 30$~$cm separation between the region of radiofrequency energy deposition by a single loop antenna and the region of maximum magnetic field applied by a pair of coils. With 200$~$W of rf input power, up to a five-fold increase in axial plasma density between the antenna and the magnetic mirror throat is observed together with an Ar II blue-mode.
Two dimensional B-dot probe measurements show that the rf magnetic fields are closely guided by the converging-diverging geometry. The wave is characterised as a $m=0$ mode satisfying the helicon dispersion relation on-axis with radial boundary conditions approximately matching the radii of the plasma column. Analysis of the wave phase velocity and wave axial damping failed to identify collisionless or collisional wave-plasma coupling mechanisms. Instead, the wave axial amplitude variations can be explained by local wave resonances and possible reflections from localised rapid changes of the refractive index. A Venturi-like effect owing to the funnel-shaped magnetoplasma and conservation of the wave energy may also explain some level of amplitude variations.
\end{abstract}

\vspace{2pc}

\submitto{\PSST}

\ioptwocol

\section{Introduction}\label{sec:Intro}

The high degree of plasma ionisation associated with helicon modes is desirable for a variety of applications, from space plasma propulsion to power generation, semiconductor manufacturing and fundamental plasma physics \cite{charles2009plasmas,agnello2019negative,kurth2017juno,chen2015helicon}. The coupling mechanisms of helicon waves have been debated since their first association with efficient high density plasma generation ($\geq 10^{12}~\rm cm^{-3}$) in the 1960's, and this question remains a dynamic field of research today both experimentally and numerically \cite{boswellthesis,boswell1984very,caneses2016collisional,lau2021helicon,chang2022first,jimenez2022wave,magarotto20203d}. 

Because electromagnetic (EM) waves of frequencies lower than the electron plasma frequency can not propagate in an unmagnetised plasma, most of the energy transfer and deposition via rf EM fields between the antenna and the plasma generally occur in a skin-layer region close to the antenna \cite{guittienne2013generation,chabert2011physics}. These non-penetrating capacitive / inductive stochastic and ohmic heating processes make it challenging to promote plasma generation in the bulk, a desirable feature in numerous aforementioned applications. In magnetised plasmas on the other-hand, while the capacitive and inductive contributions can persist, helicon waves can penetrate deep within the plasma where they can couple to the bulk electrons to contribute to the plasma generation \cite{jimenez2022wave}.

The well-known helicon dispersion relation shows that in general the wavelength is proportional to the ratio of the applied magnetic field intensity $B_0$ over the plasma density \cite{klozenberg1965dispersion}. It also appears that discharges in which the contribution of helicon waves have been verified were operating in regimes for which several wavelengths could fit within the plasma characteristic dimensions \cite{degeling1996plasma,guittienne2013generation,caneses2016collisional,guittienne2021helicon}. These and other studies have observed the collisionless and/or collisional helicon waves contributions to plasma generation in two general categories of operating conditions.

The first contribution occurs for moderately high densities ($10^{11}-10^{12}\rm cm^{-3}$) for which the dominant coupling mechanisms appear to be collisionless \cite{degeling1996plasma,guittienne2013generation}. At the radiofrequencies commonly used, these densities require applied magnetic fields where $B_0 < 100~$G in order to fit several wavelengths within the discharge. Under these conditions, it was found that the plasma density is maximised when the helicon wave phase velocity is of the order of the electron thermal speed and/or the speed of electrons most likely to ionise, e.g. $1-3\times 10^{8}~\rm cm ~s^{-1}$ for a 3$~$eV Maxwellian electron population \cite{chen1991plasma,degeling1996plasma}. In this wave mode, the plasma density is seen to increase as $\propto P_{\rm rf}^{x}$ with $x > 1$ \cite{degeling1996plasma}. This is likely owing to electrons being trapped and accelerated by the helicon waves axial electric field \cite{chen1991plasma,degeling1996plasma,lafleur2010plasma}.

The second category of discharges typically concerns densities $\sim 10^{13}~\rm cm^{-3}$ in which electron-neutral and electron-ion collisions alone can explain most of wave energy deposition \cite{light1995axial,caneses2016collisional}. At these densities, higher magnetic fields of $B_0>300~$G can be used without excessively increasing the wavelength, and the higher densities could also partially result from reduced plasma-wall losses. Moreover, in favourable experimental configurations, the beating of standing helicon waves could also increase the wave-electron coupling \cite{chi1999resonant,guittienne2021helicon,takahashi2016standing,mingyang2022relationship}. Finally, electron inertia effects resulting in electrostatic (ES) contributions - known as Trivelpiece-Gould (TG) modes - are often regarded as yet another power deposition channel of bounded whistler modes \cite{chen1997generalized,borg1998power,blackwell2002evidence}. However, with TG mode wavelengths typically in the order of a few millimetres in plasma of densities $\geq 10^{12}~\rm cm^{-3}$, the waves are radially highly damped and challenging to measure in the plasma centre \cite{chen1997generalized,degeling2004transitions}. Electron inertia effects are therefore not expected to contribute to the rf power deposition within the core of high-density plasmas but their role in edge antenna-plasma coupling mechanisms are still being investigated \cite{chen2015helicon,jimenez2022wave}.

The observation of a blue-core or step density changes are often interpreted as signs of wave-driven modes. However, observations seem to indicate that these features are neither sufficient nor necessary conditions to wave-heated regimes \cite{bennet2019non,chang2022first}. In particular, an experiment using a double-saddle antenna at 13.56$~$MHz in a strong magnetic mirror configuration has generated blue-core plasmas of densities $\geq 10^{12}~\rm cm^{-3}$ at moderate rf powers $\in ]200,500]~$W while no helicon waves were detected \cite{bennet2019non}. This motivated conducting wave measurements with a B-dot probe in an experiment reproducing all parameters of the former experiment except for the type of rf antenna and operating frequency, i.e. using a single-loop antenna at 27.12$~$MHz. While the new experiment closely reproduced the plasma parameters obtained in the former \cite{filleul2021characterization}, rf magnetic waves identified as $m=0$ helicon modes were detected with the B-dot probe. These measurements are presented and characterised here.

The main purpose of this study is to identify the nature of the rf magnetic waves with known wave theory and to do a preliminary analysis of the potential wave contributions to the plasma generation. The work is organised as follows; the background of bounded whistler waves is summarised in \Secref{Theo} and the experimental apparatus and diagnostics are described in \Secref{Apparatus}. The experimental data are presented and analysed in the scope of cold plasma wave theory in \Secref{Results}. Possible coupling mechanisms and explanations for  are considered in \Secref{Discussion}.

\section{Background}\label{sec:Theo}

The purpose of this section is to summarise notions of plasma wave theory which are necessary to interpret the data of this study.

\subsection{Dispersion relations}

In the limit of an homogeneous magnetised infinite and collisionless cold-plasma, the Fourier transformed wave equation for a plasma wave electric field $\mathbf E$ of angular frequency $\omega$ and wavevector $\mathbf{k}$ is \cite{stix1992waves}
\begin{equation}\label{eq:WaveEq}
	\mathbf{k} \times \left( \mathbf{k} \times \mathbf{E} \right) +\frac{\omega^2}{c^2} \large \mathbf{\bm{\varepsilon}} \cdot \mathbf{E} = \mathbf{T} \cdot \mathbf{E} = 0~,
\end{equation}
with $c$ the speed of light in vacuum and $\large \mathbf{\bm{\varepsilon}}$ the cold-plasma dielectric tensor. The general cold-plasma dispersion relation for plasma plane waves propagating at an angle $\theta$ to the applied magnetic field $\mathbf{B_0}$ is found by taking the determinant of the tensor $\mathbf{T}$ and writes 
\begin{equation}\label{eq:GeneDispRel}
	A n^4 - B n^2 + C = 0 ~,
\end{equation}
where $n = \frac{|\mathbf{k}| c}{\omega}$ is the complex index of refraction, and $A$, $B$ and $C$ are terms combining the ion and electron cyclotron frequencies ($\omega_{\rm ci}$,$\omega_{\rm ce}$), the plasma frequencies ($\omega_{\rm pi}$,$\omega_{\rm pe}$) as well as sine and cosine of $\theta$ \cite{stix1992waves}. In what follows, $k = |\mathbf{k}|$ is the modulus of the wavevector, i.e. the wavenumber.
Neglecting the ion mass and restricting the wave frequency to  $\omega_{ci} \ll \omega \leq \omega_{ce} \ll \omega_{pe}$, \Eqref{GeneDispRel} reduces to the dispersion relation of whistler waves \cite{boswell1984effect}
\begin{equation}\label{eq:WhistlerWK}
	\frac{k^2 c^2}{\omega^2} = \frac{\omega_{\rm pe}^2}{\omega\left(\omega_{\rm ce}\cos{\theta} - \omega - i\nu_{\rm eff} \right)}~,
\end{equation}
where $\nu_{\rm eff}$ is the effective electron collision frequency, included here for completeness.
When $\omega << \omega_{\rm ce} \cos{\theta}$ and in the collisionless limit ($\nu_{\rm eff}  / \omega  << 1$), \Eqref{WhistlerWK} simplifies to
\begin{equation}\label{eq:HeliconWK}
	\frac{k^2 c^2}{\omega^2} = \frac{\omega_{\rm pe}^2}{\omega\omega_{\rm ce}\cos{\theta}} \Longleftrightarrow k_{\parallel} k = \frac{e n_{\rm e} \omega \mu_{0}}{B_{0}}~.
\end{equation}
Here, $k_{\parallel} = k\cos{\theta}$ is the wavenumber component parallel to $\mathbf{B_{0}}$ (while $k_{\perp} = k\sin{\theta}$), $e$ is the elementary charge, $\mu_{0}$ the vacuum permeability and $n_{\rm e}$ the electron density. From \Eqref{HeliconWK}, it can be seen that in this limit the electron inertia is not taken into account.
If $k_{\perp} >> k_{\parallel}$, the wave collision absorption length $\delta_{\parallel}$ resulting from the imaginary part of \Eqref{HeliconWK} modified to include collisions writes \cite{chabert2011physics}
\begin{equation}\label{eq:Damping}
	\delta_{\parallel} = \frac{\omega_{\rm ce}}{k_{\perp} \nu_{\rm eff}} ~.
\end{equation}

\begin{figure}[!h]
\begin{center}
\includegraphics[width=7.25cm]{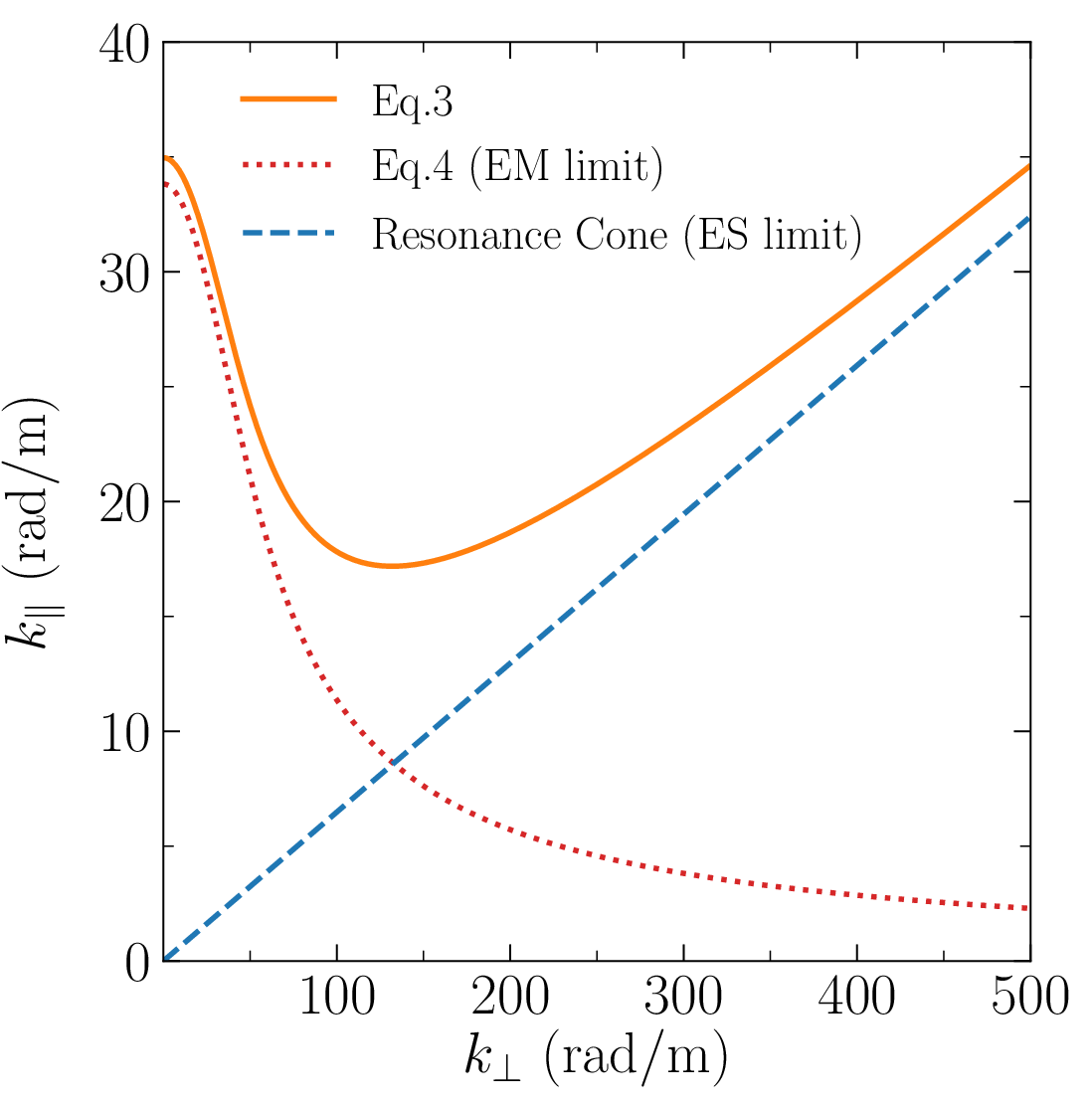}
\end{center}
\caption{The dispersion relations of \Eqref{WhistlerWK} (continuous line), \Eqref{HeliconWK} (dotted line) and the electrostatic limit (dashed line). The conditions are $B_0=150~$G, $n_{\rm e}=5\times 10^{11}~\rm cm^{-3}$ and $f=27.12~$MHz, for which $\omega/\omega_{\rm ce} \simeq 0.06$.}
\label{fig:FigDispRel}
\end{figure}

\Figref{FigDispRel} shows \Eqref{WhistlerWK} and \Eqref{HeliconWK} for conditions $B_0=150~$G, $n_{\rm e}=5\times 10^{11}~\rm cm^{-3}$ and a $27.12~$MHz wave ($\omega / \omega_{\rm ce} \simeq 0.06$). The whistler dispersion relation \Eqref{WhistlerWK} has two branches, one electromagnetic and one electrostatic, for small and large values of $k_{\perp}$, respectively. For small $k_{\perp}$, \Eqref{HeliconWK} approximates well the EM branch of \Eqref{WhistlerWK} and the wave is electromagnetic in nature \cite{borg1998power}. For large $k_{\perp}$, the electron inertia effects need to be taken into account and \Eqref{WhistlerWK} asymptotically approaches a straight line characterised by the angle $\cos{\theta_{\rm res}}=\omega / \omega_{\rm ce}$ for which the whistler wave refractive index goes to infinity \cite{degeling2004transitions}. This angle is known as the phase velocity resonance angle beyond which the whistler wave is evanescent. Whistler modes are therefore bound to propagate within a resonance cone whose main axis is along $\mathbf{B_{0}}$ and half-angle is $\theta_{\rm res}$. For $\theta$ approaching $\theta_{\rm res}$ (typically when $\omega > 0.5\omega_{\rm ce}$), the electrostatic contribution dominates and corresponds to the electron cyclotron wave \cite{degeling2004transitions}.

Since this study treats of bounded non-uniform magnetised plasmas, the infinite plane wave concepts introduced so far are expected to only provide an approximate quantification of the measured waves' properties. A model describing the inhomogeneous plasma density and magnetic field, such as performed numerically in \cite{jimenez2022wave,magarotto20203d}, is beyond the scope of this work. Taking into account the effects of the limited spatial extent of the plasma can however greatly improve the model.

\subsection{Boundary conditions}

With the inclusion of boundary conditions, the two free-space whistler dispersion relation branches develop into cavity eigenmodes, namely helicon modes and Trivelpiece-Gould (TG) modes \cite{klozenberg1965dispersion,boswell1984effect,borg1998power}. \EQREF{HeliconWK} is often called the helicon dispersion relation when used in bounded plasmas. TG waves can be understood as the result of constructive interferences of reflections of the whistler wave resonance cones from the boundaries (plasma - vacuum interface or physical surfaces) and propagate on the edge of the plasma \cite{borg1998power}.

Some studies have found that for bounded helicon waves, $k_{\parallel}$ can take discrete values determined by the cylindrical antenna length \cite{chabert2011physics,boswell1984very}, while others have observed a continuous $k_{\parallel}$ spectrum \cite{guittienne2021helicon}. In systems with conductive axial boundary conditions, step density increases have been observed for constant $B_0$ and increasing rf power, owing to resonant cavity modes resulting in discrete values of $k_{\parallel}$ proportional to the system's characteristic dimensions \cite{chi1999resonant}. Since a single loop antenna with an axial extent much shorter than the expected axial wavelength is used in this study with a plasma discharge without conductive axial boundaries, $k_{\parallel}$ is expected to exhibit a continuous spectrum.

From Maxwell's law and for fields of the form $\sim \exp{\left(i(k z - \omega t + m\theta) \right)}$, the magnetic field components of the helicon wave $\mathbf{B}$ in a uniform plasma can be expressed as combinations of Bessel functions \cite{davies1970helicon}. For example, the component of $\mathbf{B}$ parallel to $\mathbf{B_0}$ is proportional to $J_m(r k_{\perp})$, with $J_m$ the $m^{th}$ order Bessel function of the first kind, and $m$ the helicon wave azimuthal mode number. With such expressions of the $\mathbf{B}$ components, satisfying the radial boundary conditions on the edge of an insulating cylinder of radius $r=r_0$ gives
\begin{equation}\label{eq:BndCond}
	m k J_{m}(k_{\perp} r_0) + k_{\parallel} J_{m}^{'}(k_{\perp} r_0) = 0~,
\end{equation}
from which $k_{\perp}$ can be deduced depending on the azimuthal mode number, e.g. $k_{\perp}=3.83/r_0$ for the $m=0$ mode. Therefore, with $r_0$, $n_e$ and $B_0$, \Eqref{HeliconWK} and \Eqref{BndCond} can be used to compute $k_{\parallel}$ and $k_{\perp}$ of the helicon wave for a given azimuthal mode $m$. When electron inertia can be neglected, \Eqref{BndCond} holds true for both conducting and insulating boundary conditions \cite{chen1991plasma,lafleur2010plasma}. The helicon axial wavelength and phase velocity can then be calculated to be $\lambda_{\parallel} = 2\pi / k_{\parallel}$ and $v_{\phi} = f \lambda_{\parallel}$, respectively (with $f = \omega / 2\pi$). The group velocity $v_{\rm g} = \partial \omega / \partial \mathbf{k}$ can be calculated from the dispersion relations.

\section{Experimental arrangement and diagnostics}\label{sec:Apparatus}

\subsection{Apparatus}

\begin{figure*}
\begin{center}
	\includegraphics[width=14.0cm]{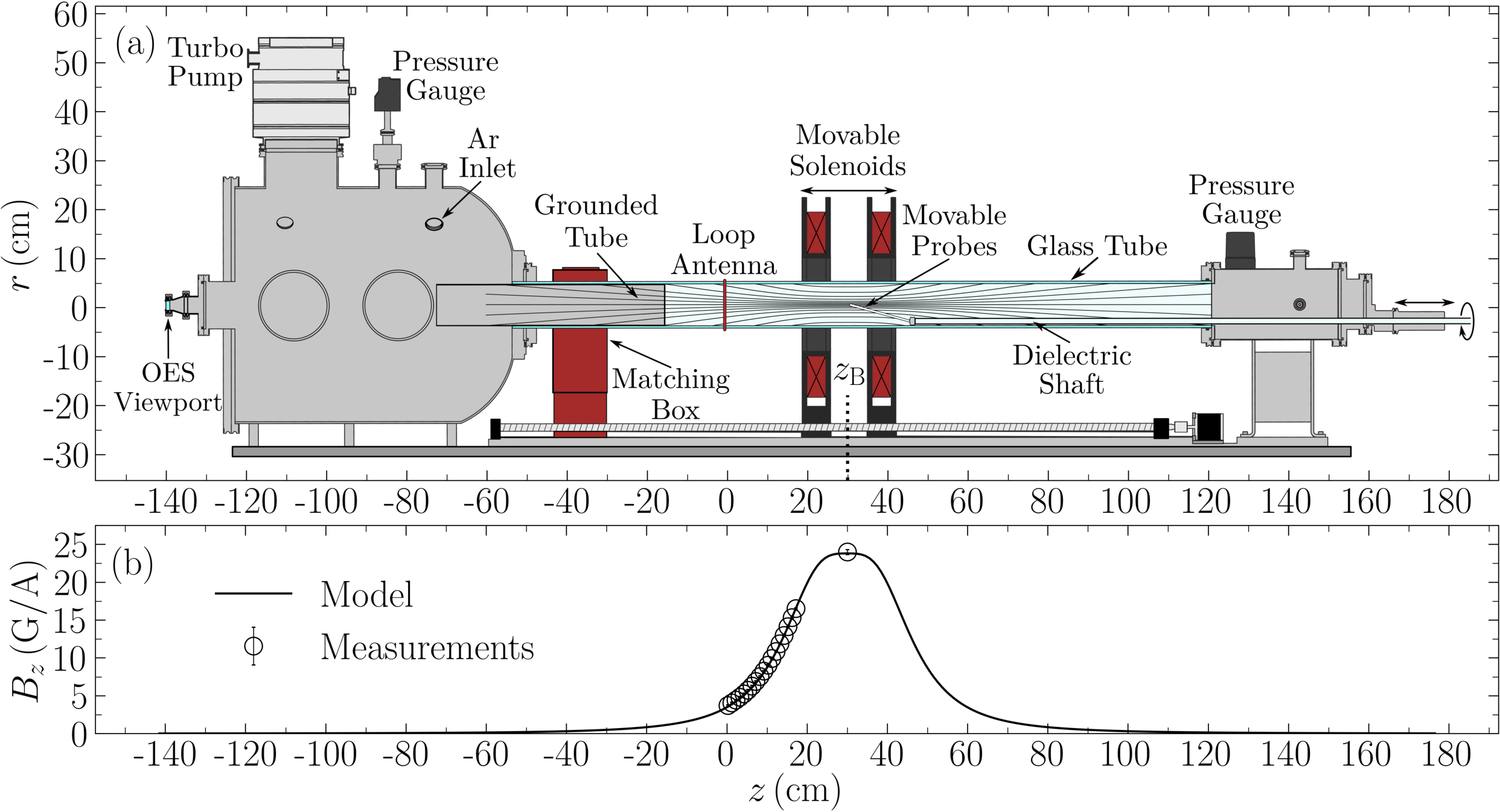}
\end{center} 
\caption{Sketch of the experimental apparatus in the configuration used in this study, i.e. with $z_{\rm B}=30~$cm and the 1-1/3 turn loop antenna at $z=0~$cm (a). Magnetic field strength on-axis $B_0$ measured with an Hall effect probe (circles) and compared with an analytical model (curve) (b).}
\label{fig:SketchPlusField}
\end{figure*}

The measurements were performed in a linear plasma device illustrated in \figref{SketchPlusField}. The active region of the apparatus consists of a $150~$cm long, $9~$cm inner diameter borosilicate glass tube connected to stainless steel vacuum chambers onto which vacuum pumps and gauges are installed to reach a base pressure of $\sim 10^{-7}~$Torr. Argon is injected from the chamber onto which the turbo pump inlet is mounted in order to  minimise axial neutral pressure gradients \cite{filleul2021characterization}. The argon working pressure ranging from 0.1 to 10$~$mTorr is set with a mass-flow controller.

The rf antenna is a 1.3 turns loop antenna made from a 3$~$mm thick copper wire wound around the glass tube. Radiofrequency power is delivered from a variable frequency rf generator through an L-type matching network made from two high-voltage variable vacuum capacitors with ranges 25$~$pF - 2000$~$pF and 10$~$pF - 500$~$pF each. The matching network ensures that the impedance matching between the 50$~\Omega$ output of the rf generator and the variable complex impedance of the load results in maximising the power transferred to the load \cite{lieberman2005principles,godyak2021rf}. The load consists of the transmission line, connectors, matching network, antenna, and the plasma. The forward and reflected power are given by the rf generator. The working frequency is set to $27.12~$MHz and the rf reflected power kept $\leq 1\%$ at all times. The antenna's centre marks the origin of the ($r$,$\phi$,$z$) laboratory reference frame used throughout this work. 

\begin{figure*}
\begin{center}
\includegraphics[width=14.cm]{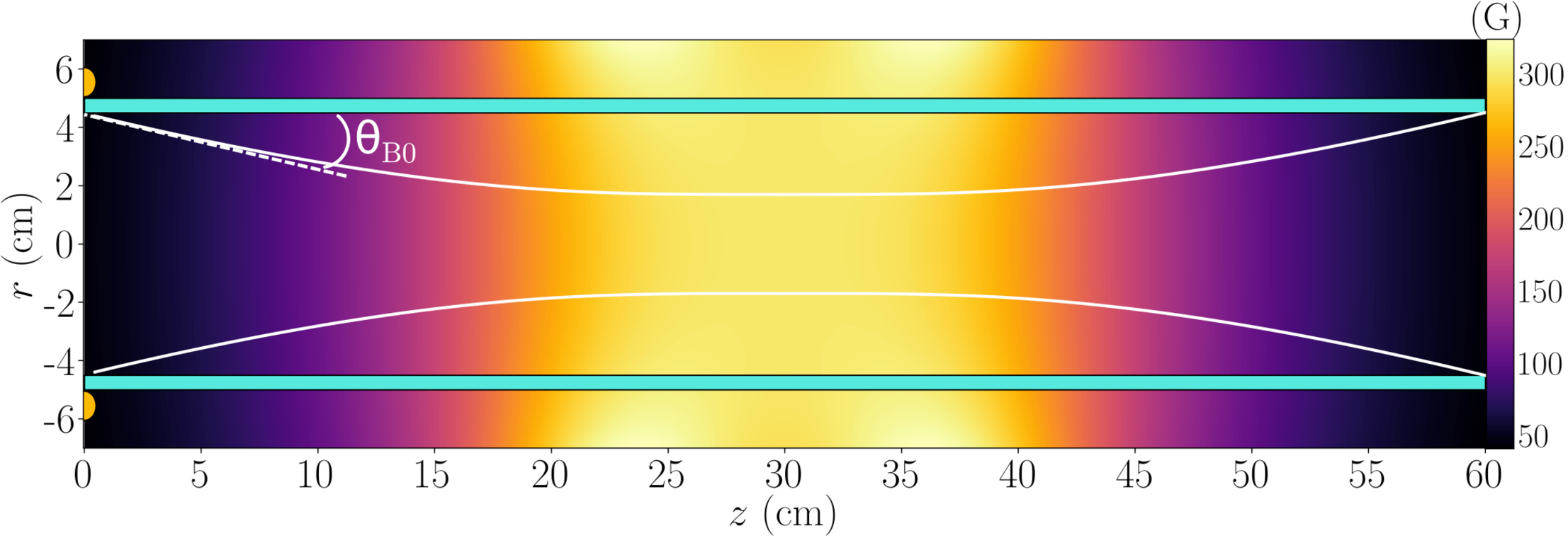}
\end{center} 
\caption{2D contours of $\mathbf{B_0}$ such that $B_0=300~$G at $(r,z) = (0,30)~$cm. The two continuous lines are the most radial magnetic streamlines to cross the antenna (orange half-dots) and to intersect the glass tube (turquoise horizontal lines) at $z=0~$cm. $\theta_{\rm B_0}$ is defined as the angle between $\mathbf{B_0}$ and $\mathbf{\hat{z}}$.}
\label{fig:B0Map}
\end{figure*}

The magnetic field is applied by a pair of movable Helmholtz coils separated by 14$~$cm centre-to-centre and placed concentrically around the glass tube. The solenoids position $z_{\rm B}$ is kept fixed, such as to place the magnetic mirror throat $30~$cm away from the antenna. This value of $z_{\rm B}$ was chosen to allow for better comparison with previous investigations which used the same configuration \cite{bennet2019non,filleul2021characterization}. The solenoids produce a peak magnetic field strength $B_{\rm 0}$ of $25~$G/A on-axis (see \figref[b]{SketchPlusField}) setting a magnetic mirror ratio of $R_{\rm m} = 6.86$ from the antenna location to the throat, as shown in \figref{B0Map}. A magnetic surface of interest is the funnel-shaped one delimited by the most radial streamlines to intersect the antenna plane (the white continuous lines in \figref{B0Map}).

\subsection{Plasma diagnostics}

The in-situ probes used in this study are mounted at the extremity of a 1.5$~$m 1/4'' steel shaft encapsulated in a glass tube in order to ensure the continuity of the dielectric plasma boundary \cite{filleul2021characterization}. This arrangement is show in \figref[a]{SketchPlusField}. The shaft slides at the bottom of the glass tube and can be rotated around $\phi$ in order for the probe tips to effectively reach every location inside the apparatus, owing to its axial symmetry. The uncertainty on the axial position of the probe is $\pm 1~$mm while the $\pm 3^{\circ}$ uncertainty on the azimuthal position equals an average error on the probe's radial coordinate of $\pm 1.8~$mm.

A planar Langmuir probe and an rf compensated cylindrical Langmuir probe (LP) have been used to estimate the ion density from the ion saturation method and the electron temperature from the Druyvesteyn method, as previously described in detail \cite{filleul2021characterization,caldarellifilleul2022data}. The effective electron temperatures $T_{\rm e}$ on-axis had only been measured at a few axial loci for the $B_0=300~$G and $B_0=600~$G cases as well as under the solenoids for other values of $B_0$. $T_{\rm e}$ on-axis was found to be varying within the range $T_{\rm e}\simeq 4.75\pm 0.5~$eV for the different magnetic fields with the $0.5~$eV error being the standard deviation around the axial mean. The error resulting from using this approximate value of $T_{\rm e}$ is propagated together with the standard deviations of the repeated ion saturation measurements to estimate the uncertainties on the ion densities reported in this study.

The B-dot probe is made out of 6 loops of $0.2~$mm copper wire forming a single coil of 4$~$mm diameter \cite{franck2002magnetic}. The coil is mounted as the extremity of a ceramic probe holder such as to measure time-varying magnetic fields along the $\mathbf{\hat{z}}$ direction. A 6$~$mm outer diameter borosilicate glass enclosure is placed around the coil to protect it from direct plasma exposure. The coil's twisted pair leads then run along the probe shaft which acts as a coaxial shield. The leads are connected to an hybrid combiner which suppresses common-mode signals associated with electrostatic pick-ups and preserves the differential magnetic signal  \cite{borg1987guided}. The common-mode rejection of the hybrid coupler was tested and found to be close to 98$\%$, i.e. when a $27.12~$MHz common-mode signal is picked up, $\sim 2\%$ of the signal leaked as a differential signal. This error, combined with the uncertainties on the ($r,z$) coordinates of the B-dot probe, was used to compute the uncertainties on the measured waves amplitudes and phases at 27.12$~$MHz. A second B-dot and hybrid combiner are used to measure the rf field of the loop antenna in atmosphere to provide a phase reference for the mobile B-dot's signal. The outputs of the hybrid combiners are recorded and digitised by a 200$~$MHz bandwidth oscilloscope and Fast-Fourier Transform (FFT) post-processed to filter out harmonics and extract the $27.12~$MHz magnetic rf wave amplitude and phase.

Argon I and II emissions at $750.4~$nm and $488~$nm are measured with a previously described arrangement made out of two 10$~$nm narrow-band-pass filters and a calibrated CMOS sensor \cite{filleul2022ion}. A feature of optical emission spectroscopy (OES) is that the intensity of emission lines can be interpreted in terms of the density and temperature of the particles contributing to their excitation. Within the density range of interest, the emission rate coefficients can be determined from the so-called corona model from the convolution of the excitation collisions cross-sections with the electron energy distribution function \cite{fantz2006basics}. Ar I emission at $750.4~$nm is excited from an electron-neutral impact from ground state and its intensity can be modelled as
\begin{equation}
	I_{\rm 750nm} \propto K_{\rm 750nm}(T_{\rm e})n_{\rm e} n_{\rm g}~,
\end{equation}
where $K_{\rm 750nm}$ is the emission rate coefficient and $n_{\rm g}$ the neutral argon density \cite{vincent2022high}. The $488~$nm Ar II line is preferably excited from an electron-ion interaction \cite{blackwell1997two}
\begin{equation}\label{eq:I488}
	I_{\rm{488nm}} \propto K_{\rm{488nm}}(T_{\rm{e}})n_{\rm e} n_{\rm i}~,
\end{equation}
with $K_{\rm 488nm}$ the corresponding emission rate coefficient. From quasi-neutrality, it follows that $I_{\rm 488nm} \propto K_{\rm{488nm}}(T_{\rm{e}}) n_{\rm e}^2$. With $4.75\pm 0.5~$eV Maxwellian energy distributions and the cross-section of direct excitation from ion ground state \cite{imre1972cross}, $K_{\rm 488nm}=1.37 \pm 0.49\times 10^{10} \rm cm^{3}/s$. This uncertainty on $K_{\rm 488nm}$ can be reported on the values of $n_{\rm i}$ inferred from \Eqref{I488}. In this study the filtered CMOS sensor was placed on a viewport at $z=-140~$cm such that the recorded intensities are integrated along the plasma column axial line of sight.

Finally, an \textit{Octiv} $I$-$V$ probe in-line between the matching box and the loop antenna is used to measure the plasma resistance $R_{\rm p}$ from the current, voltage and respective phase at the antenna terminal. The circuit resistance $R_{\rm ant} = 0.34\pm0.01~\Omega$ was measured when operating the apparatus with no plasma to allow calculation of the power coupling efficiency $\eta$
\begin{equation}\label{eq:eta}
	\eta = \frac{R_{\rm p}}{R_{\rm ant} + R_{\rm p}}~.
\end{equation}
The uncertainty on $\eta$ is computed from propagating the measurements errors on $R_{\rm ant}$, on the applied power, the \textit{Octiv} probe outputs, as well as by correcting for the impact of the piece of transmission line between the output of the \textit{Octiv} probe and the antenna.

\section{Results}\label{sec:Results}

For continuity with previous studies conducted in similar apparatuses, the rf power, argon pressure and solenoids positions were kept fixed at 200$~$W, 1$~$mTorr ($0.13~$Pa) and $z_{\rm B} = 30~$cm, respectively throughout this work. Only the applied magnetic field intensity $B_0$ was varied and its impact on the plasma and rf magnetic waves characterised. The properties of the plasma discharge are summarised here to allow the comparison between the measured waves features and the dispersion relations. For now, the plasma is assumed to be a medium carrying the waves and the focus is on the characterisation of the waves propagating across the converging-diverging magnetic field. Possible wave-plasma coupling mechanisms will be discussed in the following section.

\subsection{Plasma discharge characteristics}

\begin{figure}[!h]
\begin{center}
\includegraphics[width=7.25cm]{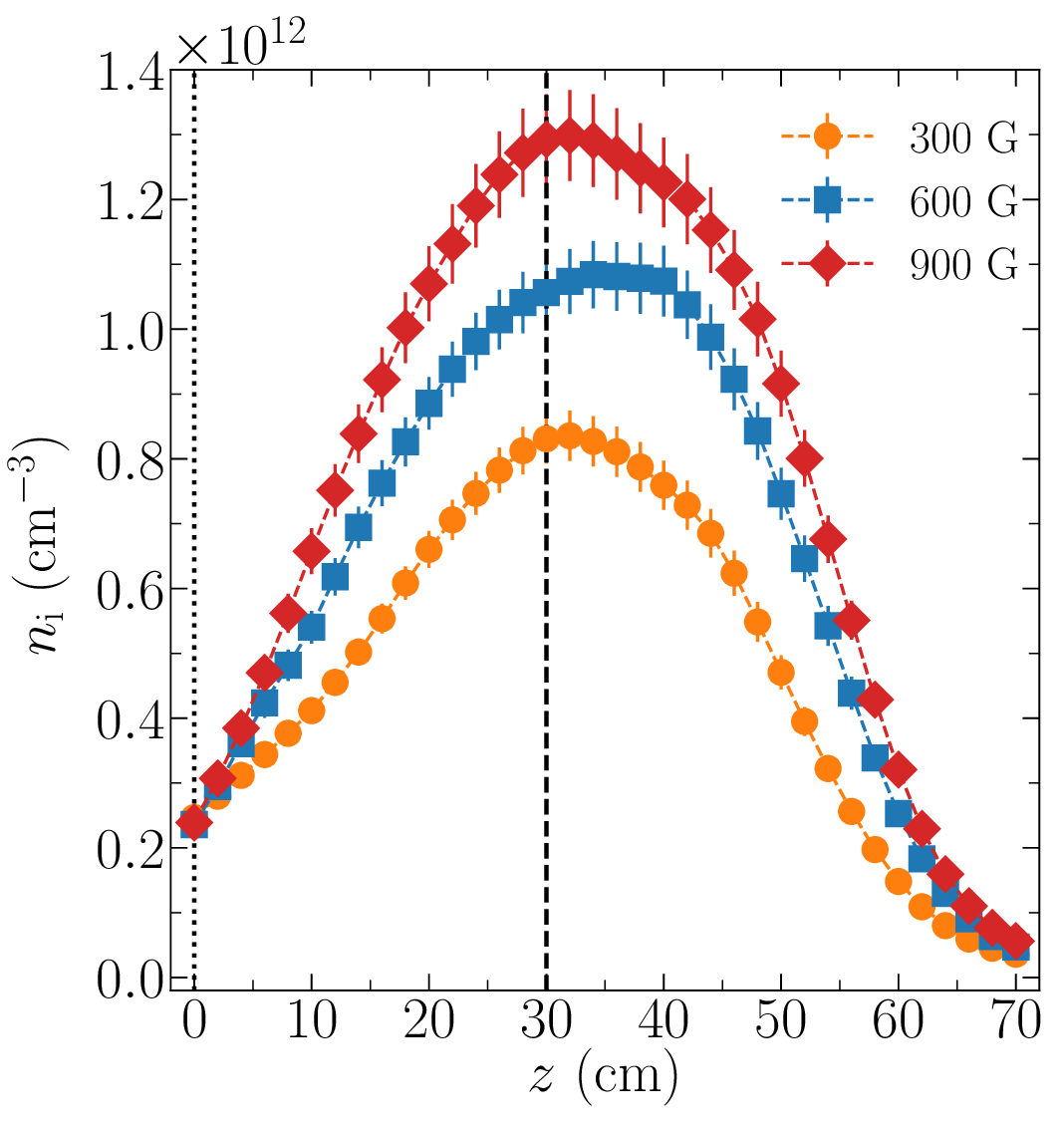}
\end{center}
\caption{Axial plasma density measured with the LP for increasing $B_0$. The dashed vertical line marks the position of the centre of the solenoids and the dotted one the position of the antenna. The error bars are the uncertainties on $n_{\rm i}$ as detailed in \Secref{Apparatus}.}
\label{fig:PlasmaDensIncB0Prf}
\end{figure}

The plasma obtained under such experimental conditions was first characterised in \cite{bennet2019non} with an apparatus of identical dimensions and configuration to the one used in this study, but employing a double-saddle antenna at 13.56$~$MHz. The antenna - magnetic mirror separation of $z_{\rm B} = 30~$cm was found to maximise the plasma generation under the solenoids in the converging-diverging plasma column when $B_0$ is greater than a threshold value of 250$~$G. For $B_0$ below this threshold, the maximum achieved density is reduced and the axial density profile is bimodal with one maximum under the antenna and one under the solenoids. It was shown that these features and the plasma parameters absolute values were closely maintained when changing the antenna and radiofrequency to a single-loop antenna driven at 27.12$~$MHz \cite{filleul2021characterization,filleul2022ion}.

\begin{figure}[!h]
\begin{center}
\includegraphics[width=7.cm]{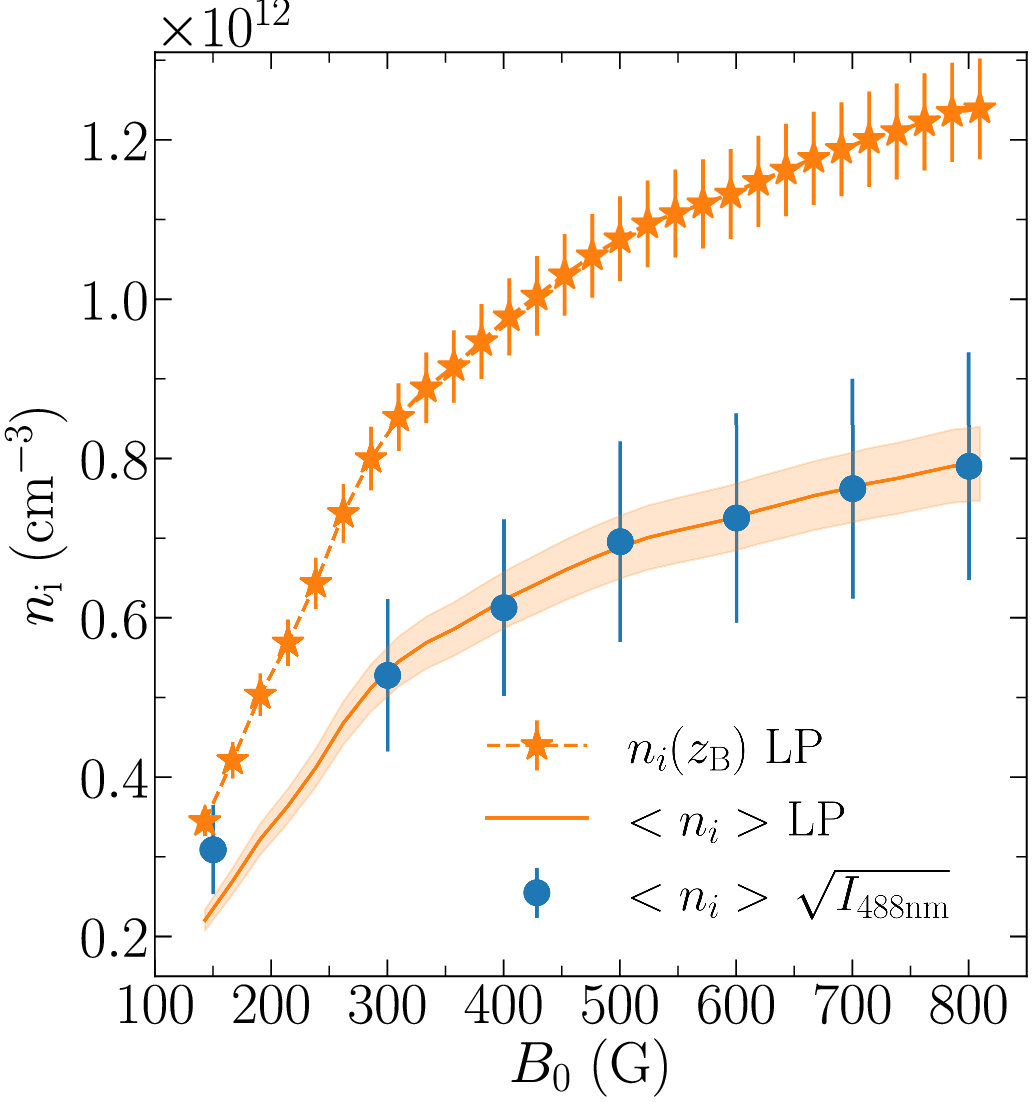}
\end{center}	
\caption{Ion density under the solenoids on-axis ($z=30~$cm) measured with the LP (star markers) and the corresponding axially averaged density (continuous line) for increasing $B_0$. The square-root of the axially integrated Ar II line emission (blue dot markers) is scaled to the axially averaged $n_{\rm i}$ at 300$~$G.}
\label{fig:PlasmaDensIncB0}
\end{figure}

\Figref{PlasmaDensIncB0Prf} shows the single peaked axial ion density profiles for $B_0=300~$G, 600$~$G and 900$~$G. The profiles are roughly symmetrical with respect to the solenoids (dashed vertical line) and the maximum density monotonously increases under the solenoids despite remaining approximately constant under the antenna ($z=0~$cm, vertical dotted line). As previously observed, this increase under the solenoids is well explained by the reduction of plasma losses to the walls as the maximum density approaches the value expected from the geometrical reduction of the magnetic field cross-section (see \figref{B0Map}) \cite{bennet2019non,filleul2021characterization}. This convergence can be further noted in \figref{PlasmaDensIncB0} from the curve showing the ion density measured with the Langmuir probe under the solenoids at $z=30~$cm for increasing $B_0$. The effective electron temperature measured on-axis under the solenoids was observed to be approximately unchanged at $T_{\rm e} ~\simeq5.2 \pm 0.2~$eV between the three $B_0$ cases of \figref{PlasmaDensIncB0Prf}.

\begin{figure}[!h]
\begin{center}
\includegraphics[width=8.25cm]{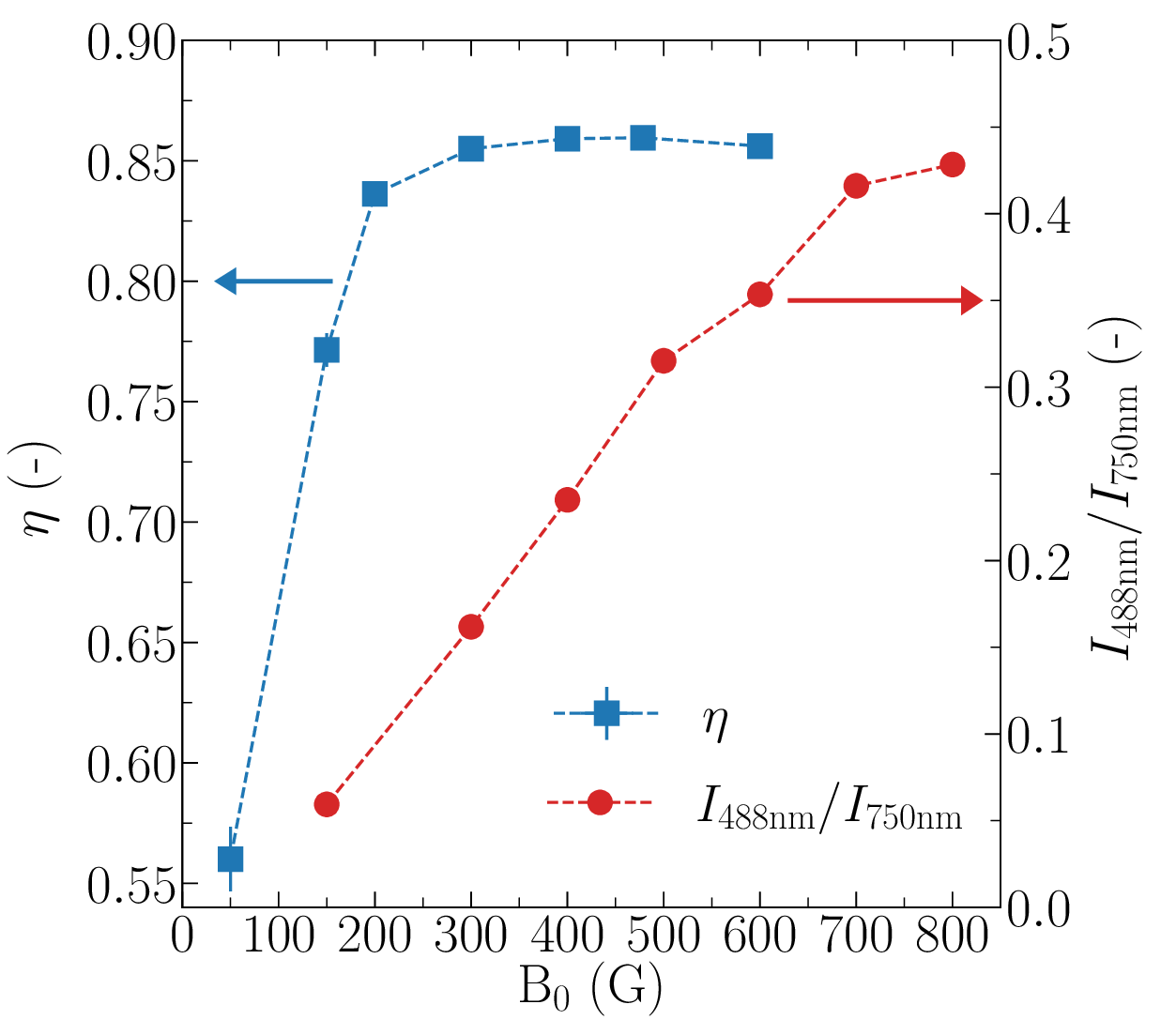}
\end{center} 
\caption{Power transfer efficiency $\eta$ (left-axis) and ratio of Ar II over Ar I intensities along the centreline of the plasma column (right-axis) for increasing $B_0$.}
\label{fig:ArIIArIRatioEta}
\end{figure}

In what follow, the wave features are analysed from both the local and macroscopic (axially averaged) perspectives. Since the medium is non-uniform, the measured macroscopic axial wavelengths are best interpreted with dispersion relations using the axially averaged plasma densities and magnetic fields. Unfortunately, the axial plasma density profiles were only measured with the LP for the $B_0$ cases in \figref{PlasmaDensIncB0Prf}. To remedy to this, the axially integrated Ar II intensities $I_{\rm 488nm}$ were recorded for all considered values of $B_0$. From \Eqref{I488}, taking isothermal electrons, the Ar II intensity is quadratically proportional to $n_{\rm i}$. As such, the square-root of $I_{\rm 488nm}$ should be a good qualitative estimate of $<n_{\rm i}>$, the plasma density axial average. The absolute value of $<n_{\rm i}>$ at 300$~$G is computed from \figref{PlasmaDensIncB0Prf} and compared to the respective value of $\sqrt{I_{\rm 488nm}}$ to obtain a scaling factor. This factor is applied to all $B_0$ cases and the optically determined values of $<n_{\rm i}>$ are plotted in \figref{PlasmaDensIncB0} as a function of $B_0$ (dot markers). The uncertainties on the optically deduced $<n_{\rm i}>$ are deduced from the uncertainty on the $K_{\rm 488nm}$ emission rate coefficient.

To validate this approach, the trend of $<n_{\rm i}>$ $\sqrt{I_{\rm 488nm}}$ for increasing $B_0$ is compared with the same trend of $<n_{\rm i}>$ estimated from the Langmuir probe data only. The LP $<n_{\rm i}>$ values for all $B_0$ cases are computed by scaling the on-axis LP measured density $n_{\rm i}$ at $z=30~$cm from \figref{PlasmaDensIncB0} (star markers) to the axial average values $<n_{\rm i}>$ calculated from the three $B_0$ cases of \figref{PlasmaDensIncB0Prf}. This produces the continuous line in \figref{PlasmaDensIncB0}. The good agreement between the OES $<n_{\rm i}>$ and the Langmuir probe $<n_{\rm i}>$ gives confidence in the OES deduced $<n_{\rm i}>$ values.

Along with the increase in plasma density, the ratio of $I_{\rm 488nm}$ to the 750$~$nm emission line $I_{\rm 750nm}$ along the apparatus centreline is also seen to monotonously increase with $B_0$, as shown in \figref{ArIIArIRatioEta}. Assuming that the respective emission rate coefficients are varying at similar rates, the ratio $I_{\rm 488nm} / I_{\rm 750nm}$ can be taken as a proxy for the ratio $n_{\rm e} / n_{\rm g}$, as a first approximation. The trend of the ratio is not only the result of the increasing $I_{\rm 488nm}$(i.e. $n_{\rm e}$) with $B_0$, which appears to plateau for high values of $B_0$, but also of the decreasing $I_{\rm 750nm}$(i.e. $n_{\rm g}$) on-axis, likely owing to neutral depletion \cite{magee2013neutral}. Indeed, considering room temperature neutrals, the ionisation fraction under the solenoids is $\sim 2.4\%$ when $B_0=300~$G. Following equation 9.6 from \cite{chabert2011physics} which balances the neutral pressure radial profile with the centrally peaked electron pressure, the neutral density in the centre would amount to $\sim 20\%$ of the neutral density at the edge of the column. At 400$~$G, $I_{\rm 488nm}$ roughly equals a quarter of $I_{\rm 750nm}$ and a blue-mode is observed.

Finally, \figref{ArIIArIRatioEta} shows the antenna-plasma coupling efficiency $\eta$, for increasing $B_0$ calculated from \Eqref{eta}. Above $B_0=300~$G the coupling efficiency is seen to stabilise at $\sim 85\%$. This shows that beyond 300$~$G, a constant amount of power gets deposited into the plasma from the antenna and that the change in density is likely not the result of a mode transition but rather of reduced losses from improved magnetic confinement.

\subsection{Spatio-temporal wave features}

\subsubsection{Two dimensional characteristics.}

The B-dot probe was incrementally moved by 2$~$cm steps in $\mathbf{\hat{z}}$ and $10^{\circ}$ in $\mathbf{\bm{ \hat{\phi}}}$, to obtain time-resolved 2D features of $B_{\parallel}$. Due to the eccentric placement of the probe shaft, a $10^{\circ}$ increment equals a $\simeq 6~$cm step in the $\mathbf{\pm \hat{r}}$ direction. The rf magnetic pick-up coil of the B-dot probe is always oriented normal to the laboratory frame of reference's $\mathbf{\hat{z}}$ direction. The angle $\theta_{B_0}$ is defined as the angle between the converging-diverging magnetic streamlines and $\mathbf{\hat{z}}$. Within the glass-tube volume of interest, it takes a maximum values of $\theta_{B_0} = \pm 12^{\circ}$ (see \figref{B0Map}). As a result, at most $2.2\%$ of the amplitude of the rf magnetic field measured by the B-dot probe also comprises a $B_{\perp}$ component. This was corrected by multiplying the measured signals by the local value of $\cos{\theta_{B_0}}$. On-axis, $B_{\parallel}$ is purely along $\mathbf{\hat{z}}$. The following data analysis is performed on the FFT-extracted $27.12~$MHz fundamental frequency component of the raw signal. The first and second harmonic components are not expected to play a significant role in the wave behaviour as their respective amplitudes were always lower than $10\%$ of the fundamental frequency amplitude.

\begin{figure*}
\begin{center}
\includegraphics[width=14.5cm]{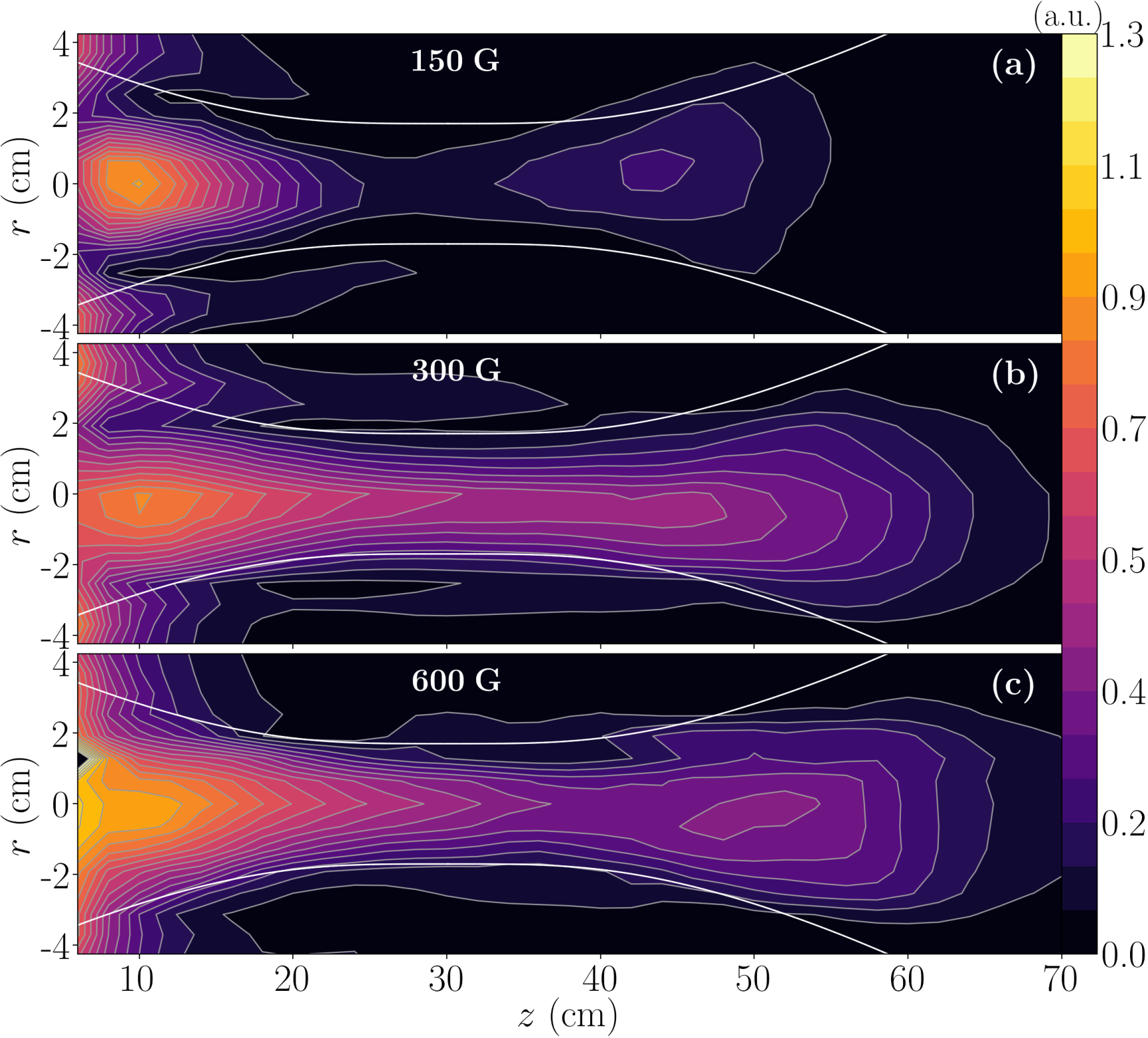}
\end{center} 
\caption{Two dimensional profiles of $|B_{\parallel}|$ for $B_0=150~$G (a), $B_0=300~$G (b) and $B_0=600~$G (c). The white lines show the most radial streamlines.}
\label{fig:BzAmp2D}
\end{figure*}

The 2D measurements were carried out for three magnetic field intensities: $B_0=150~$G, 300$~$G and 600$~$G. The resulting maximum amplitudes of $B_{\parallel}$ are show in \figref{BzAmp2D}. The region corresponding to $z<6~$cm was not plotted in order to increase the contrast of the wave features. Indeed, under the antenna, the magnitude of magnetic field induced by the rf current flowing through the antenna dominates the rf EM wave field excited elsewhere. For the three values of $B_0$, the magnetic component of the EM rf waves were picked-up by the B-dot probe down to $z=60~$cm and $z=70~$cm. The first wave feature of interest is the waveguide-like effect of the magnetised plasma on the rf waves. This is probably mainly the result of the well-known ability of whistler waves to propagate along the magnetic streamlines. Outside the funnel-shaped volume delimited by the most radial streamlines, the plasma density being lower than within the funnel, the appropriate wave dispersion relation might not be locally satisfied. As such the wave might be evanescent outside the funnel-shaped volume, explaining the wave amplitude nearing zero at these loci in \figref{BzAmp2D}. Possible explanations of the rf wave intensity variations between the $B_0=150~$G case on one end and the 300$~$G and 600$~$G cases on the other end are discussed in \Secref{Discussion}.

\begin{figure*}
\begin{center}
\includegraphics[width=14.5cm]{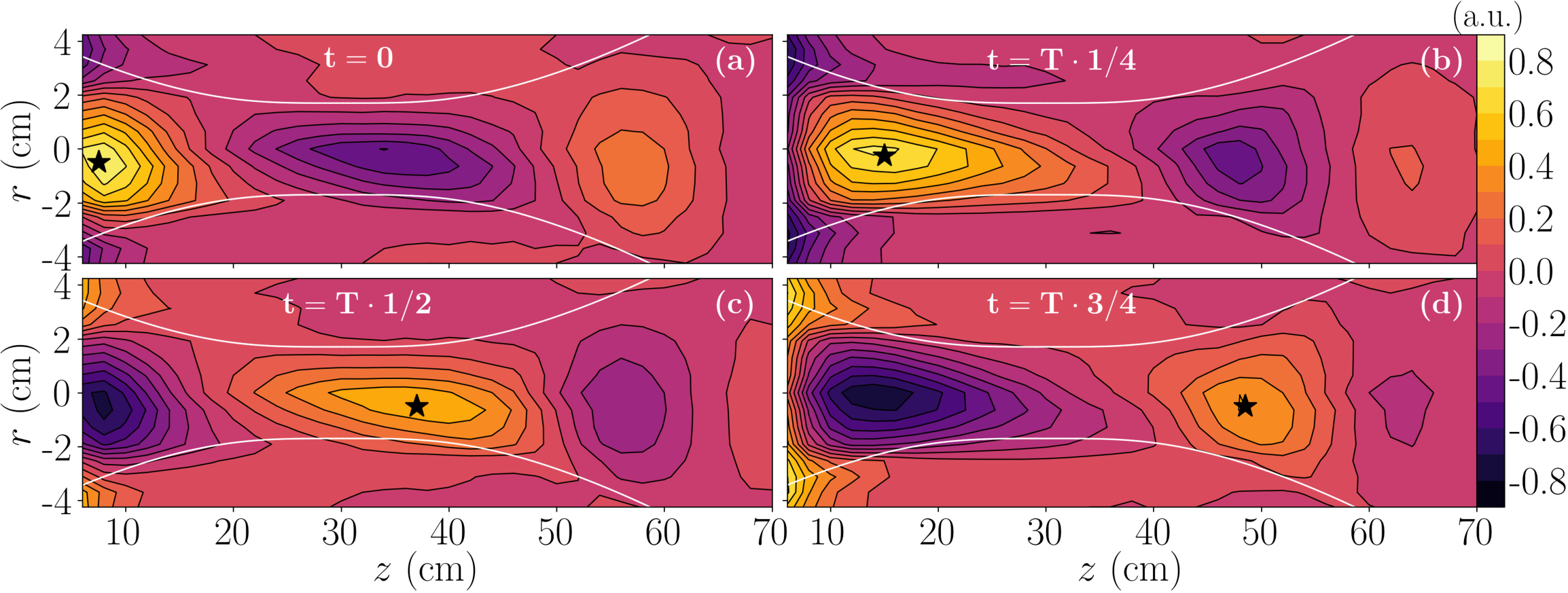}
\end{center} 
\caption{Temporal evolution of $B_{\parallel}$ for four quarters of an rf period $T$ (a-d) illustrating the travelling-wave behaviour of $B_z$ at the conditions of \figref[b]{BzAmp2D}, i.e. when $B_0 = 300~$G. The black star serves as a guide to follow one wavefront.}
\label{fig:2DTravellingWave}
\end{figure*}

The second interesting feature in \figref{BzAmp2D} is that the overall appearance of the 2D profiles of $B_{\parallel}$ closely resembles one of an $m=0$ helicon mode \cite{degeling2004transitions,boswell1984very}. In particular, the radial profiles of $B_{\parallel}$ resemble Bessel functions of the first kind $J_{0}(k_{\perp}r)$. From the boundary conditions \Eqref{BndCond} for an $m=0$ azimuthal mode, it can be found that $k_{\perp}=3.83/r_0$, where $r_0$ is the radial coordinate of the plasma boundary \cite{lieberman2005principles}. As such, $r_0$ is used as a free parameter in the fitting of $J_{0}(k_{\perp}r)$ to the radial profiles of $B_{\parallel}$. From the best fit at each $z$, axial profiles of $k_{\perp}$ can be obtained and will be used later.

An example of fitting $J_{0}(k_{\perp}r)$ to the data is shown in \figref[a]{RadialBzAmp} for the $B_0=300~$G case at $z=18~$cm and the good agreement between $J_{0}(k_{\perp}r)$ and the data is observed. The characteristic $m=0$ mode side-lobes are present and resolved from $z=8~$cm to $z\simeq 38~$cm in this case. For $B_0=150~$G, while the side-lobes feature is pronounced close to the antenna, they are quickly damped together with the central peak around $z=30~$cm, likely owing to the waves becoming evanescent, as later detailed in \figref{VphRes}. For the $B_0=600~$G case, the fitting parameter $r_0$ is greater than the glass tube radius for $z<15~$cm (see \figref[b]{RadialBzAmp}) and the side-lobes would develop outside the plasma column and are therefore not observed.

The values of $r_0$ giving the best fit are consistent with the funnel shape of the wave profiles, as it can be seen in \figref[b]{RadialBzAmp} for the three cases shown in \figref{BzAmp2D}. Focussing on the $B_0 = 300~$G case, $r_0$ is roughly symmetrical with respect to the solenoids up to $z\simeq 50~$cm, beyond which it increases to values greater than the glass-tube inner radius ($4.5~$cm). It could be that this behaviour is the result of the wave boundary conditions changing from being a radial density gradient between $z=10~$cm and $z=50~$cm, to being the dielectric surface of the glass-tube elsewhere.

\begin{figure}[!h]
\begin{center}
\includegraphics[width=7.cm]{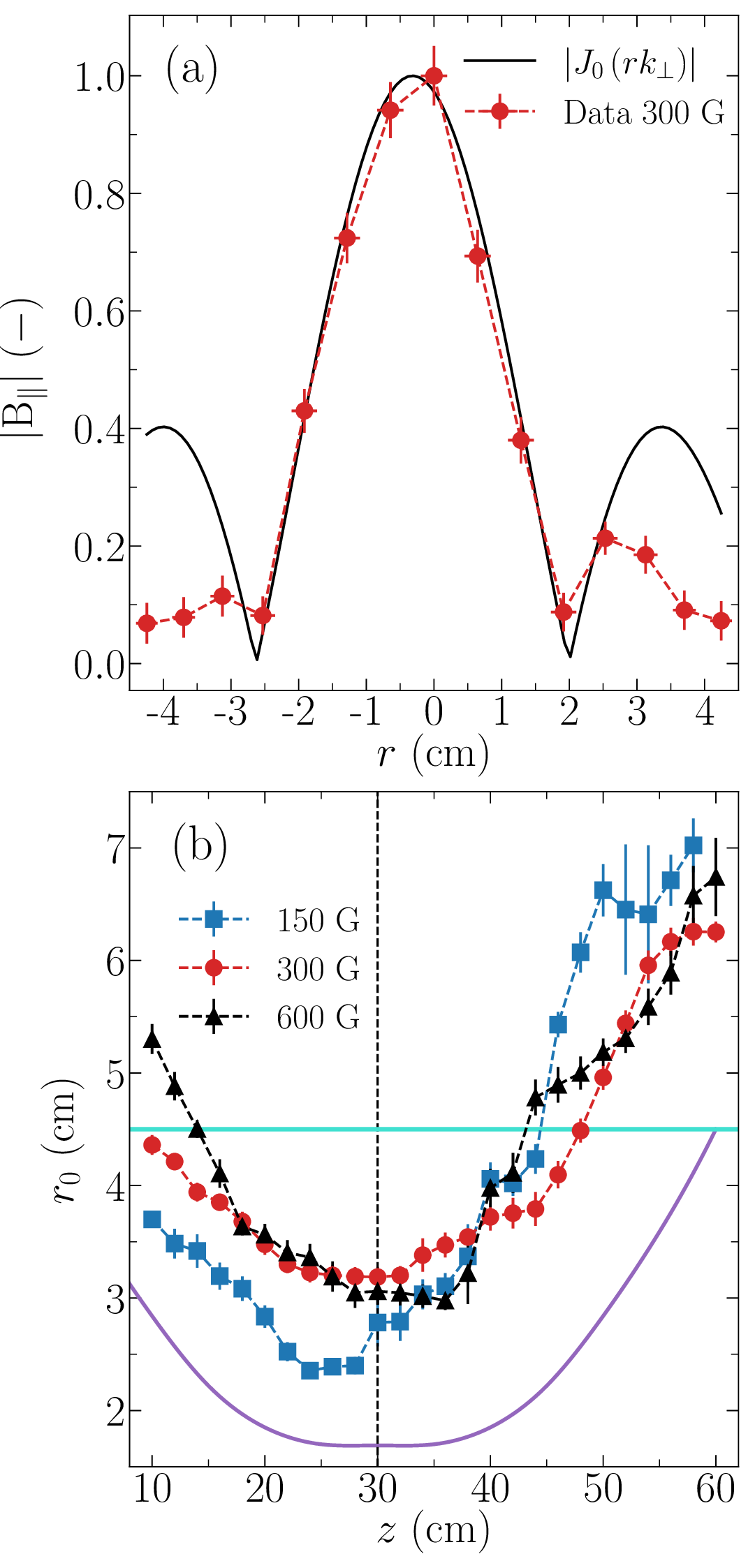}
\end{center}
\caption{Normalised $|B_{\parallel}|$ radial profiles taken from \figref[b]{BzAmp2D} ($B_0=300~$G) at $z=18~$cm (dot markers) with the fitted Bessel function (a). Fitting parameter $r_0$ versus $z$ for $B_0=150~$G (square markers), $B_0=300~$G (dot markers) and $B_0=600~$G (triangle markers), together with the radial coordinates of the most radial streamline (continuous curve) (b). The horizontal line represents the inner surface of the glass-tube.}
\label{fig:RadialBzAmp}
\end{figure}

\Figref{2DTravellingWave} shows the travelling wave feature of the measured $B_{\parallel}$ for $B_0=300~$G by displaying the spatial variation of $B_{\parallel}$ at four intervals of an rf period. The wavefront can be followed moving along $\mathbf{\hat{z}}$. From the distance separating two crests, a parallel wavelength of approximately 50$~$cm can be estimated. An animated version of \figref{2DTravellingWave} is available as supplementary material.

\subsubsection{Wave axial behaviour.}

The evolution of $|B_{\parallel}|$ on-axis for increasing values of $B_0$ is shown in \figref[a]{WaveFeatIncB0}. From the measured exponential decay rate of the magnetic field induced by the antenna's rf current on-axis when $B_0=0~$G, it was found that the rf EM wave amplitude dominates the antenna's induced magnetic field for $z>8~$cm whenever $B_0 \geq 150~$G. As such, the analysis of the rf waves feature are conducted from $z=10~$cm to $z=70~$cm in what follows. The $B_0=150~$G case stands out as being more strongly damped than the rest, which themselves appear to all behave similarly. The axial plasma density for $B_0=150~$G being bimodal and of overall much lower amplitude than the profiles for $B_0 \geq 300~$G (see \figref{PlasmaDensIncB0}) is thought to be the cause for the difference in wave damping and will be discussed below. The roughly equal wave amplitude profiles for increasing $B_0$ would indicate that wave-plasma coupling is not significantly affected by the change of $B_0$ and the resulting expected increase in wavelength. This hints that the increase in plasma density and in Ar II emission with increasing $B_0$ might not be correlated with the presence of the wave.

\begin{figure}[!h]
\begin{center}
\includegraphics[width=7.cm]{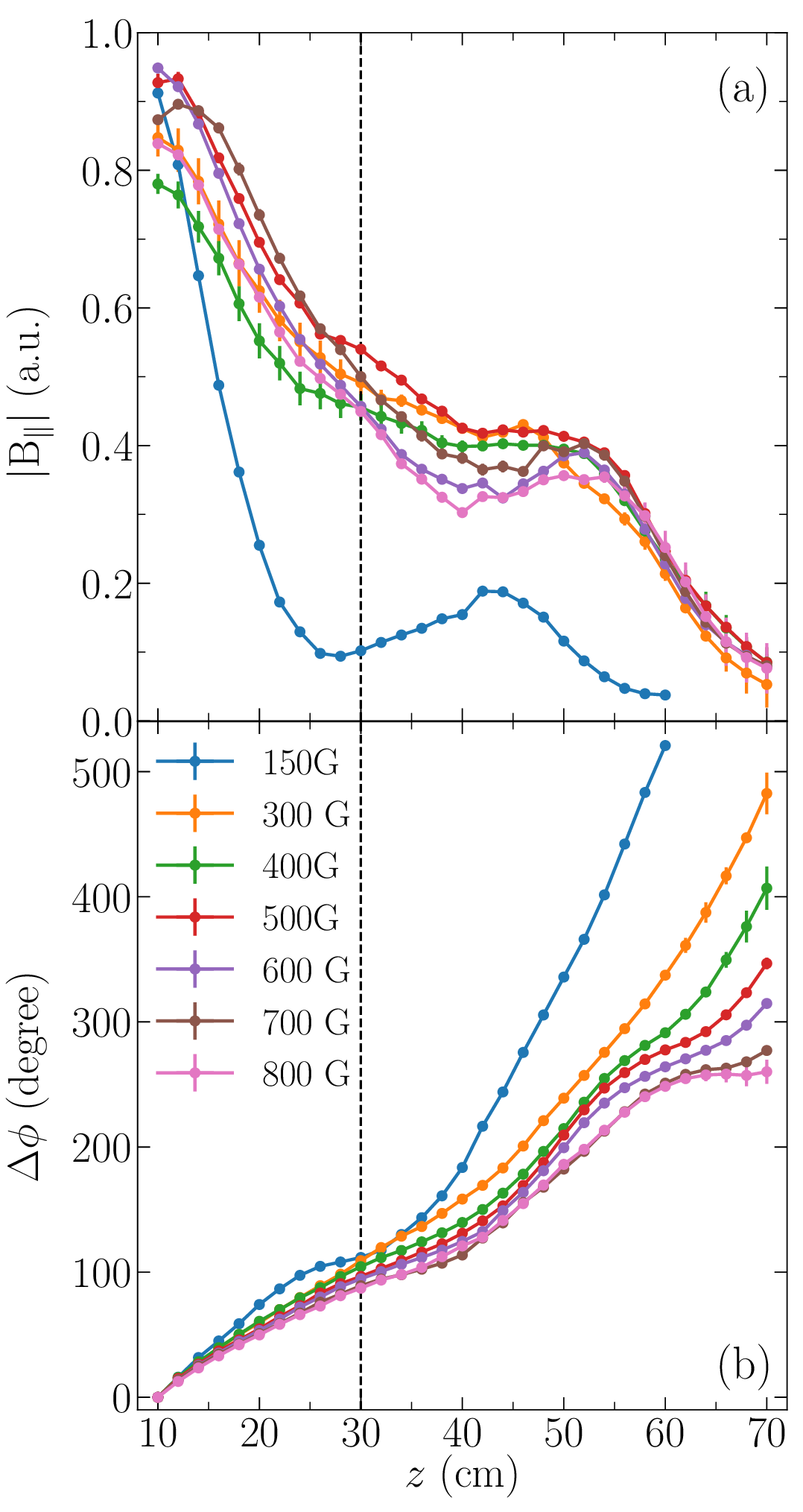}
\end{center}	
\caption{Axial amplitude (a) and phase (b) of $B_{\parallel}$ for increasing $B_0$. The error bars are deduced from the $2\%$ of the electrostatic pick-up not rejected by the hybrid combiner.}
\label{fig:WaveFeatIncB0}
\end{figure}

\Figref[b]{WaveFeatIncB0} shows the phase shift $\Delta \phi$ resulting from the wave propagating away from $z=10~$cm and obtained from the FFT processing. Standing waves profiles are usually characterised by $180^{\circ}$ phase jumps coinciding with local minima of $|B_{\parallel}|$ \cite{degeling1996plasma,chi1999resonant}. Since such features are not observed but instead the phase continuously increases; it can be therefore concluded that the measured waves have the dominant features of travelling waves. The fact that for some magnetic field intensities the wave amplitudes feature local maxima at $z\simeq 50~$cm is discussed in \Secref{Discussion}.

From \figref[b]{WaveFeatIncB0}, an estimate of the macroscopic parallel wavelength $\lambda_{\parallel}$ can be calculated from the distance it takes the phase to change by $360^{\circ}$, i.e.
\begin{equation}\label{eq:LambdaEq}
\lambda_{\parallel} = 360\frac{\Delta z}{\Delta \phi} ~.
\end{equation}
Finally, the local parallel wavenumber can be computed from $k_{\parallel} = \partial \phi / \partial z$.

\subsection{Dispersion relation}

\begin{figure}[!h]
\begin{center}
\includegraphics[width=7.cm]{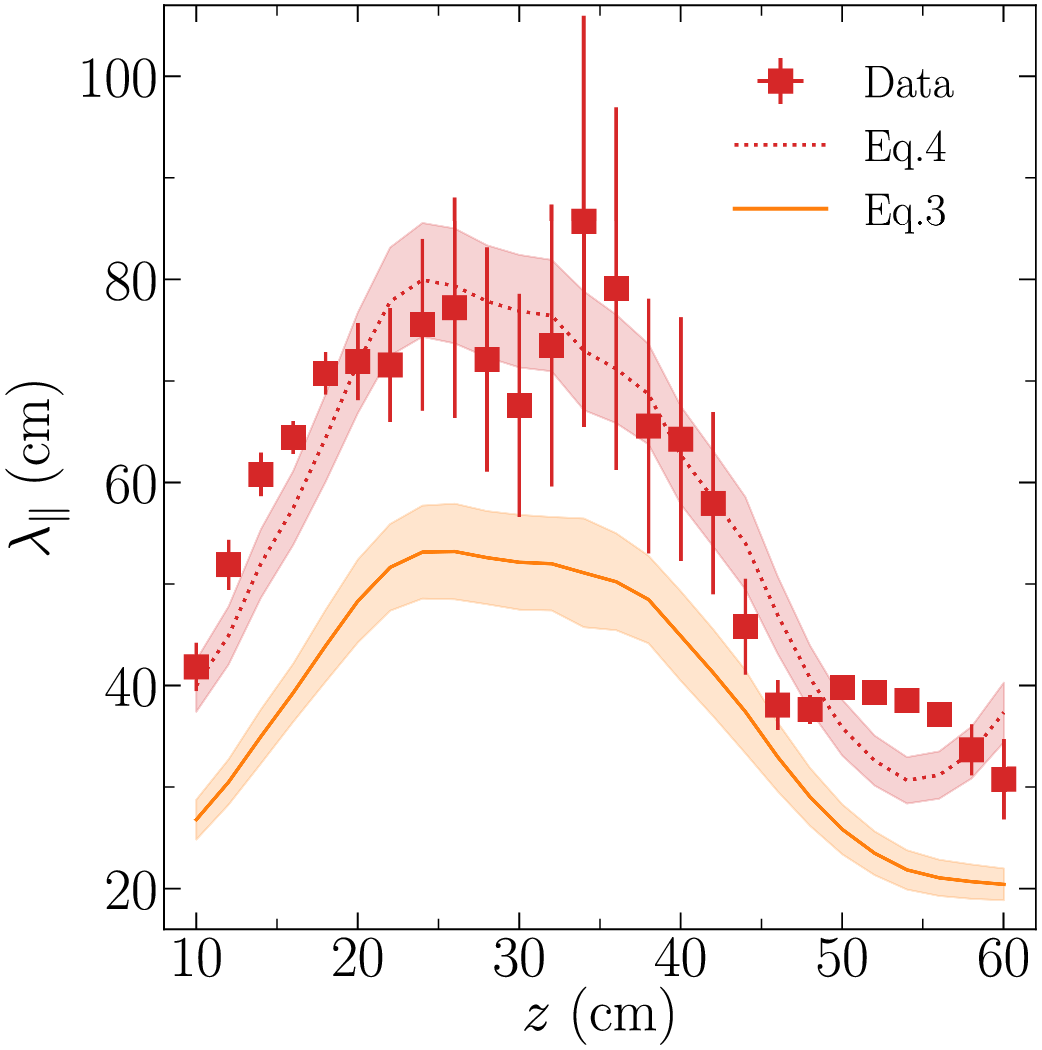}
\end{center}	
\caption{Local parallel wavelength (square markers) for the $B_0=300~$G case compared with the theoretical values of $\lambda_{\parallel}$ deduced from the helicon dispersion relation \Eqref{HeliconWK} (red dotted curve) as well as the whistler dispersion relation \Eqref{WhistlerWK} (continuous orange curve). The shaded areas represent the uncertainty on $\lambda_{\parallel}$.}
\label{fig:LocalLambda}
\end{figure}

\begin{figure}[!h]
\begin{center}
\includegraphics[width=7.5cm]{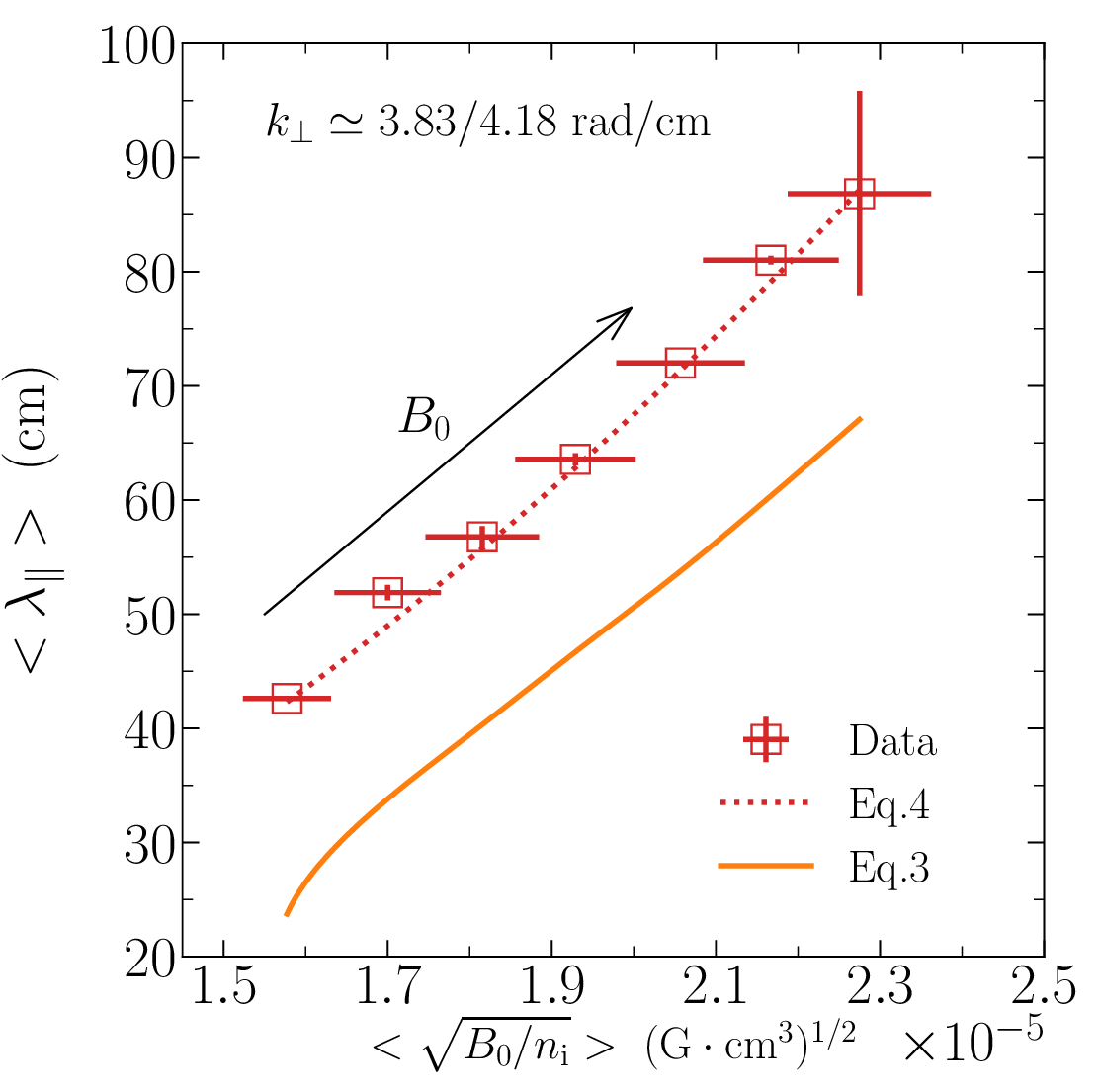}
\end{center}	
\caption{Comparison between the axially averaged measured parallel wavelength (red squares) and the helicon dispersion relation (\Eqref{HeliconWK}) for a radially bounded wave with $k_{\perp} \simeq 3.83/4.18~$rad/cm (red dotted curve). The whistler dispersion relation (\Eqref{WhistlerWK}) is also given for comparison (continuous orange curve). The abscissa axially averaged quantities are taken from \figref{PlasmaDensIncB0}.}
\label{fig:Dispersion}
\end{figure}

The local values of $\lambda_{\parallel}$, deduced from the axial derivative of the phase shift of \figref[b]{WaveFeatIncB0}, are shown in \figref{LocalLambda} for $B_0=300~$G. The error bars are obtained from uncertainty propagation and the higher noise level compared to \figref[b]{WaveFeatIncB0} results from the differentiation noise amplification. Using the axial profiles of $B_0$ and $n_{\rm i}$ from \figref{PlasmaDensIncB0Prf} with the $k_{\perp}$ values from \figref[b]{RadialBzAmp}, the theoretical values of $\lambda_{\parallel}$ can be calculated according to the helicon relation (\Eqref{HeliconWK}). The agreement from $z=10~$cm to $z= 50~$cm between the data and the helicon dispersion relation supports that the observed local wave is an $m=0$ helicon mode. The same process is repeated with \Eqref{WhistlerWK} to plot the $\lambda_{\parallel}$ values expected from the generalised dispersion relation which includes electron inertia. This solution follows the same trend as the measured vales of $\lambda_{\parallel}$ while underestimating them. This is further discussed in \figref{Dispersion}. Interestingly, from $z=10~$cm to $z= 50~$cm, $\omega \ll \omega_{\rm ce}$, validating the use of \Eqref{HeliconWK}. For $z \geq 52~$cm, however, $\omega / \omega_{\rm ce} \geq 0.1$ and the electron inertia is expected to start influencing the dispersion, possibly explaining the observed local discrepancy between the data and the helicon dispersion relation \cite{boswell1984effect,degeling2004transitions}.

\Figref{Dispersion} repeats the process of \figref{LocalLambda} for the axial macroscopic quantities of the $B_0$ cases treated in \figref{WaveFeatIncB0}. The measured values of $<\lambda_{\parallel}>$, obtained with \Eqref{LambdaEq}, increase continuously with $<\sqrt{B_0 / n_{\rm i}}>$, i.e. without steps, as expected from the used antenna and the discharge axial boundary conditions. For $B_0=300~$G, $<\lambda_{\parallel}> \simeq 52~$cm and matches the visual assessment made in \figref{2DTravellingWave} as well as the average of the local $\lambda_{\parallel} \simeq 53~$cm in \figref{LocalLambda}. The axially averaged value of $r_0$, obtained from \Eqref{BndCond} and the Bessel function fitting process shown in \figref[b]{RadialBzAmp}, gives an average value of $k_{\perp} \simeq 91.6~\rm rad/m$. With this value of $k_{\perp}$, the agreement between the helicon dispersion relation and the data is excellent, further supporting that the measured rf waves are $m=0$ helicon waves on-axis over the explored range of $B_0$. Reporting this value of $k_{\perp}$ in \figref{FigDispRel} (which used the $B_0=300~$G average conditions), it can be seen that the measured rf waves are on the helicon branch of the whistler mode and this holds true for $B_0 \geq 300~$G.

Since $k_{\perp}$ is deduced from \Eqref{BndCond} which neglects the effects of electron inertia, the potential effects of this assumption can be evaluated by using $k_{\perp} \simeq 91.6~\rm rad/m$ in \Eqref{WhistlerWK} and plotting the obtained $<\lambda_{\parallel}>$ values in \figref{Dispersion}. The constant off-set from 300$~$G to 800$~$G implies that the effect of having neglected the electron inertia is constant. For $150~$G, however, the \Eqref{WhistlerWK} curve diverges from the measured value $<\lambda_{\parallel}> \simeq 43~$cm, meaning that electron inertia effects start to play a more significant role. This coincides with $\omega / \omega_{\rm ce} \simeq 0.12$ for this condition.Therefore, in accordance with previous studies, the contributions of the electrons inertia to the wave dynamics and plasma coupling might only become significant globally when $B_0 < 150~$G and locally in regions of the discharge where $\omega / \omega_{\rm ce} > 0.1$ for $150 ~\rm G \leq B_0 \leq 600~$G \cite{degeling2004transitions,chen1997generalized,jimenez2022wave}. Resolving to more complete boundary conditions than \Eqref{BndCond} to include electron inertia would likely result in a better agreement between the data and the generalised dispersion relation. Nevertheless, given the good agreement between the data and the helicon dispersion relation, this model is adequate enough to further investigate the waves' behaviour on-axis and refining the analysis to include electron inertia effects is left for future work.

\section{Discussion}\label{sec:Discussion}

The aim of this section is to use the known properties of helicon waves in preliminary quantitative and qualitative analyses of the possible wave-plasma coupling mechanisms and the 2D wave behaviours in the inhomogeneous plasma.

\subsection{Wave-plasma coupling}

The rf magnetic wave being identified as an helicon wave, one can now check whether the damping observed on-axis is owed to wave-plasma collisionless or collisional coupling. Among other studies, \cite{chen1991plasma} and \cite{degeling1996plasma} have found strong correlations between enhanced plasma generation and the presence of helicon waves having phase velocities close to the electrons' thermal speed, or to the electrons' speed most likely to ionise, respectively \cite{lafleur2010plasma}. 

\begin{figure}[!h]
\begin{center}
\includegraphics[width=7.cm]{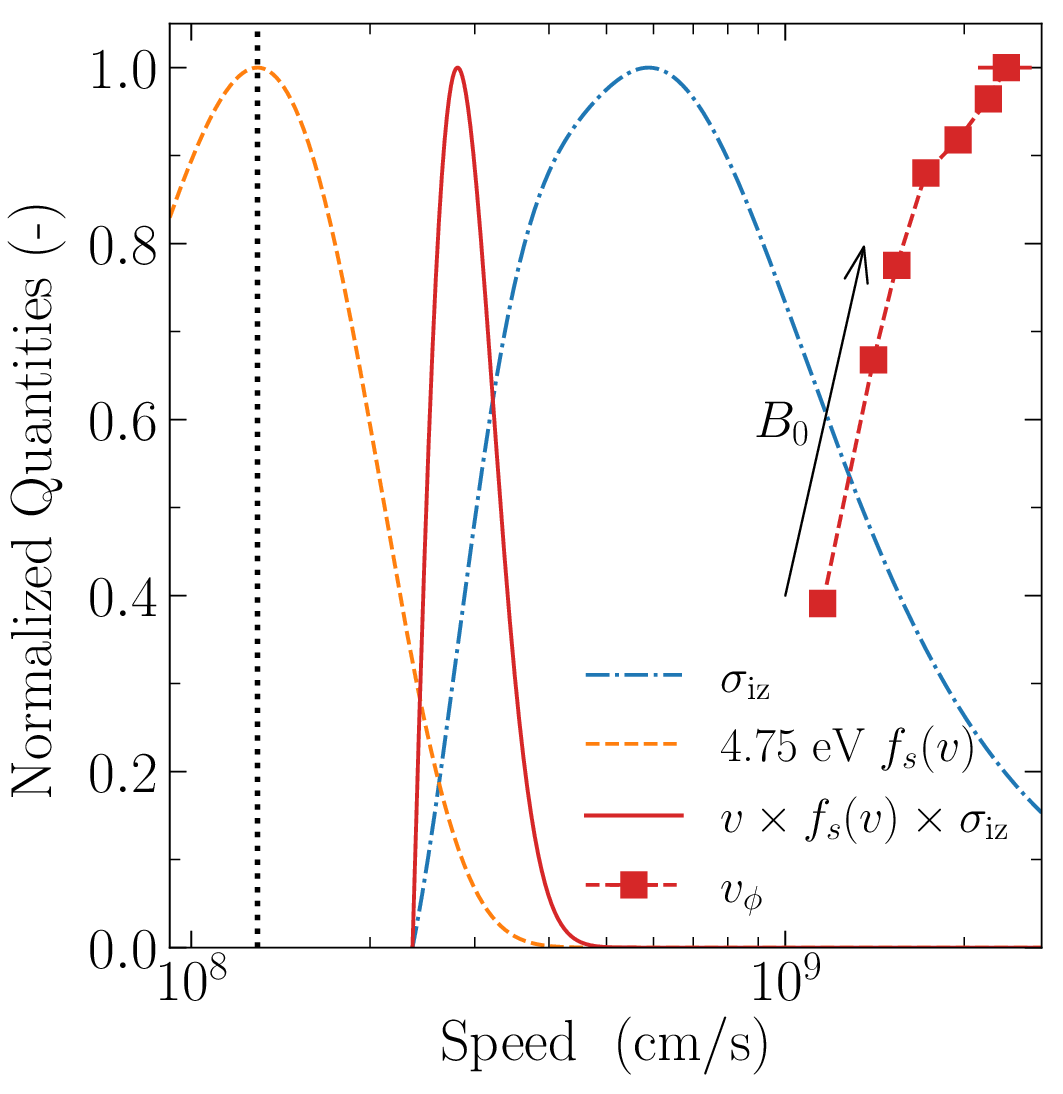}
\end{center}
\caption{Comparison between the velocities of electrons most likely to ionise for a 4.75$~$eV Maxwellian speed distribution function (continuous line), and the measured helicon wave phase velocities $v_{\phi}$ (square markers). $v_{\phi}$ is plotted against the normalised averaged plasma density as $B_0$ increases in the direction indicated by the arrow.}
\label{fig:NoWaveIonization}
\end{figure}

The phase velocities $v_{\phi}$ corresponding to the measured values of $\lambda_{\parallel}$ in \figref{Dispersion} are plotted in \figref{NoWaveIonization} for increasing $B_0$ and therefore for increasing plasma density. Also plotted is a 4.75$~$eV normalised Maxwellian speed distribution function $f_s(v)$, which is the average axial value of the effective electron temperature obtained from integrating the measured electron-energy probability functions (EEPF) for $B=300~$G and $600~$G. The vertical dotted line in \figref{NoWaveIonization} shows the distribution's thermal speed. The measured values of $v_{\phi}$ are at least one order of magnitude larger than the thermal speed, thus the wave-electron coupling mechanism detailed in \cite{chen1991plasma} can be ruled out as potential cause of the wave damping on-axis.

Given $f_s(v)$ and the argon ionisation cross-section $\sigma_{\rm iz}$, the ionisation rate constant is \cite{lieberman2005principles}
\begin{equation}\label{eq:KIZ}
	K_{\rm iz} = \int_{0}^{\infty} v f_s(v) \sigma_{\rm iz} \, dv ~.
\end{equation}
\Figref{NoWaveIonization} shows $\sigma_{\rm iz}$ values taken from \cite{phelps1999cold} alongside the integrand of \Eqref{KIZ}. The maximum of this integrand gives the electrons having the speed most likely to cause ionisation for a 4.75$~$eV Maxwellian population, i.e. $\sim 2.8\times 10^{8}~\rm cm.s^{-1}$, or $22.4~$eV. In \figref{NoWaveIonization}, discharges with higher plasma densities (higher $B_0$) yielded higher values of $v_{\phi}$. This trend took place despite $v_{\phi}$ moving further away from the speed of electrons most likely to ionise. Moreover, at energies corresponding to the measured values of $v_{\phi}$ (e.g. $> 376~$eV), the electron number densities are expected to be negligible and are unlikely to cause any quantifiable wave damping. These observations indicate that wave-trapping as described in \cite{degeling1996plasma} is unlikely to be the cause of observed wave damping on-axis.

\begin{figure}[!h]
\begin{center}
\includegraphics[width=8.cm]{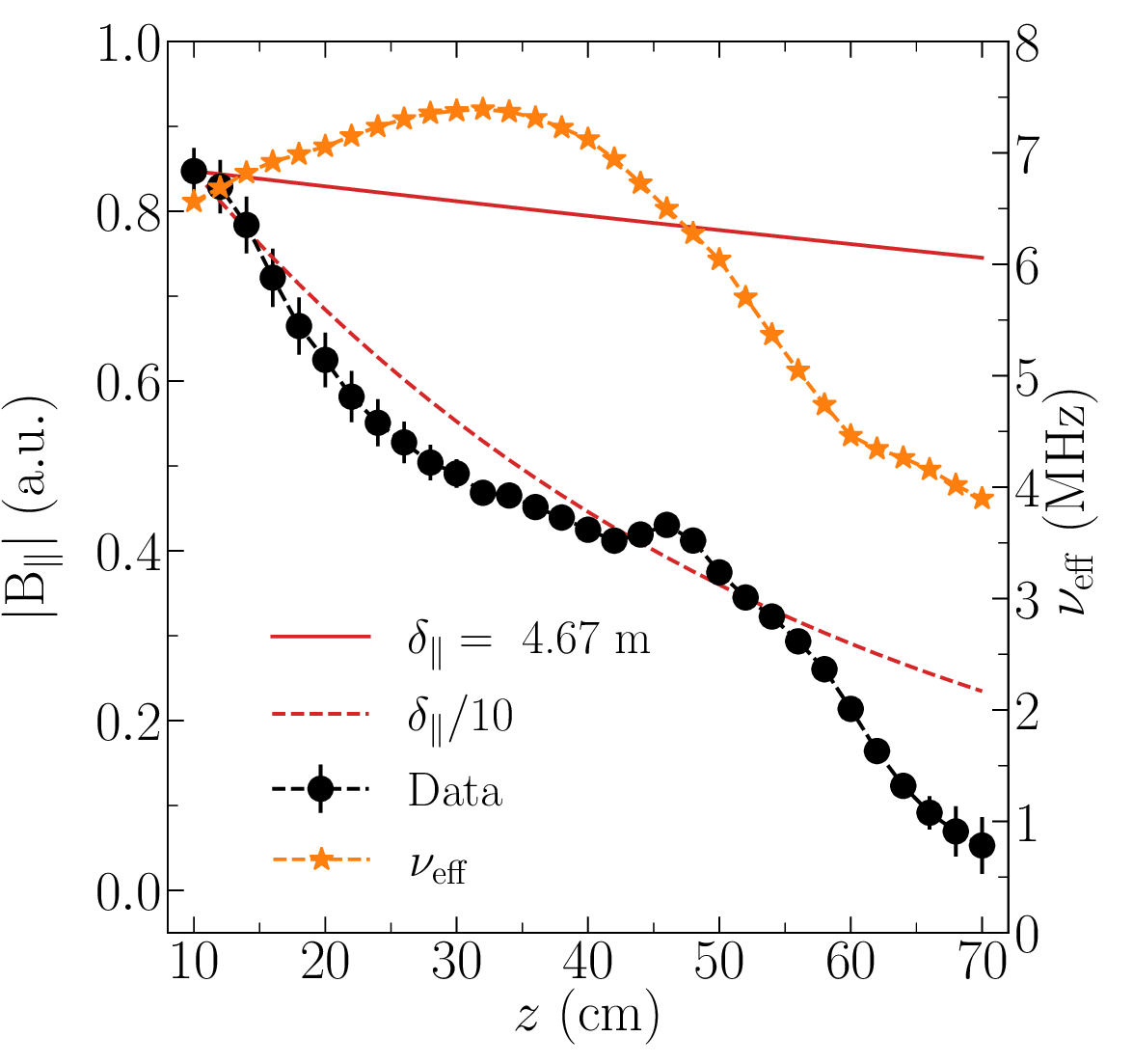}
\end{center}
\caption{$B_0=300~$G case $|B_{\parallel}|$ (dot markers) and wave damping estimated from \Eqref{Damping} (continuous curve) and with $\delta_{\parallel}/10$ (dashed line). The computed effective collision frequencies $\nu_{\rm eff}$ (star markers) are reported on the right-hand axis.}
\label{fig:Collision}
\end{figure}

With the collisionless processes likely ruled out, the wave collision absorption length $\delta_{\parallel}$ from \Eqref{Damping} can be calculated using $\nu_{\rm eff} = \nu_{\rm ei} + \nu_{\rm en}$. $\nu_{\rm ei}$ is the electron-ion Coulomb collision frequency \cite{lieberman2005principles} while $\nu_{\rm en}$ combines the elastic scattering, ionisation and excitations collisions frequencies \cite{chabert2011physics,alves2014lisbon}. $\nu_{\rm en}$ is calculated from rewriting \Eqref{KIZ} with the appropriate cross-section and multiplying the integral by the neutral density. For $B_0=300~$G, using the EEPFs measured on-axis, and not accounting for ion-pumping for simplicity, $\nu_{\rm eff}$ was calculated for each $z$ position and plotted in \figref{Collision}. The electron-neutral elastic scattering collision is the dominant process accounting for around $5~$MHz. With $\nu_{\rm eff} \sim 7~$MHz, $\nu_{\rm eff} / \omega \simeq 0.04$ and therefore collisional processes cannot account for the axial wave damping. To further illustrate this, $\delta_{\parallel}$ can be estimated to be $\sim 4.7~$m, and used with
\begin{equation}\label{eq:expdamp}
	B_{\parallel}(z) = B_{\parallel}(z_0)\exp{\left( -(z-z_0) / \delta_{\parallel} \right)}
\end{equation}
to produce the continuous curve in \figref{Collision}. This curve represents the approximate wave damping if a collisional process was the principal cause of it. Interestingly, dividing $\delta_{\parallel}$ by 10 produces the dashed curve which approximates the measured damping well. Similar observations of anomalous damping were made in \cite{boswell1984very} and spawned the quest for collisionless coupling mechanisms. In recent years, various instability-driven anomalous transports of electrons have been reported in helicon-type plasmas \cite{lee2011measurements,hepner2020wave,takahashi2022wave}. These instabilities can lead to the effective electron collision frequency to be an order of magnitude or more greater than when classical collisional processes only are considered. As such, instabilities could be a pathway for the collisional wave-plasma coupling to nevertheless explain the observed high damping rates. Here, \Eqref{expdamp} was used rather than the ray-tracing Wentzel-Kramers-Brillouin (WKB) method to estimate the overall axial wave attenuation from wave-plasma coupling in this inhomogeneous plasma. This is because the WKB method's validity condition $\left|\frac{\partial n / \partial z}{n}\right|^{-1} >> \lambda_{\parallel}$, i.e., that the wavelength needs to be much less than the gradient scale length of the refractive index, was found to be only strictly fulfilled for limited axial extents (typically $\sim 10~$cm under the solenoids), for the conditions encountered in this study \cite{stix1992waves}. Resorting to complex wave optics formalisms is beyond the scope of this study.

\subsection{Reflections and resonances}

Reflections of helicon waves at the axial conducting boundaries of plasma columns have been associated with standing wave patterns when the reflected wave interacts constructively with the forward wave and such situations seem to sometimes improve the wave-plasma coupling \cite{chi1999resonant,guittienne2021helicon}. Local variations in plasma densities from changes in the applied magnetic fields are also known to affect the waves propagation and coupling \cite{lafleur2010plasma,takahashi2016standing}. In the present apparatus, the plasma density drops by several orders of magnitude before reaching the normal conducting surface at the downstream end of the apparatus and it can be seen from \figref{BzAmp2D} that the wave amplitudes go to zero well before even exiting the glass tube. Nevertheless, for $B_0 > 150~$G, the wave axial amplitudes feature a clear local maximum at $z\simeq 50~$cm (see \figref[a]{WaveFeatIncB0}) which may be the result of wave reflections. The plasma column being highly inhomogeneous, rapid changes of the index of refraction $n$ can be expected and a fraction of the wave would be reflected and another absorbed by resonant processes \cite{stix1992waves}. Qualitatively speaking, from the lack of standing-wave features in \figref{2DTravellingWave} and \figref[b]{WaveFeatIncB0}, it can inferred that the reflected wave amplitude on-axis is negligible compared to the amplitude of the forward wave. The high axial damping rate would also act in reducing the amplitude of any wave travelling backward. Let's now consider the case of resonances.

\begin{figure*}
\begin{center}
\includegraphics[width=14.5cm]{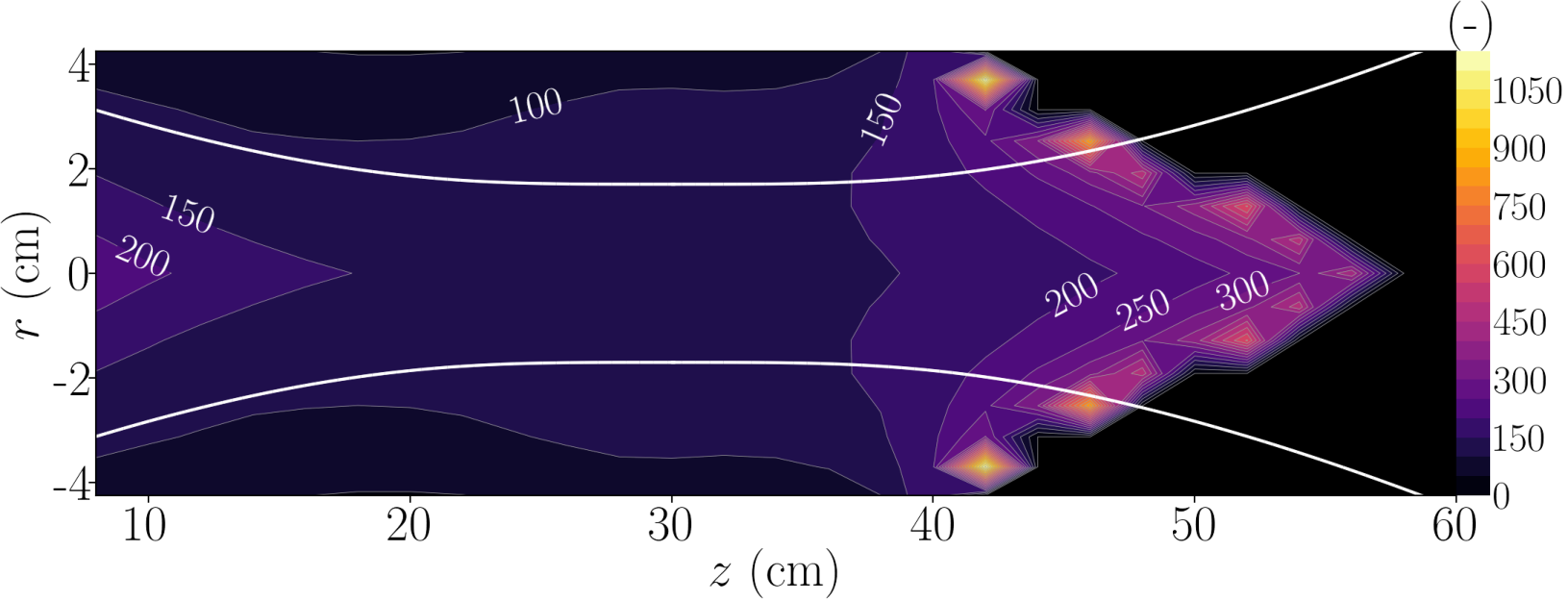}
\end{center}
\caption{Index of refraction $n$ calculated with \Eqref{GeneDispRel} using the measured volume plasma density $n_{\rm i}$ and calculated $\mathbf{B_0}$. For this example, an ersatz wave is set to travel at $\theta = 79.6^{\circ}$ with respect to the local $\mathbf{B_0}$, the axial average of \figref[b]{VphRes}. Regions in black show where $n$ is purely imaginary while it is purely real elsewhere.}
\label{fig:2DNref}
\end{figure*}

First note that electron cyclotron resonance surfaces are well outside the volume of interest for the operating conditions of interest. Following the steps in \cite{takahashi2016standing}, the local values of $n$ are calculated from \Eqref{GeneDispRel} for the $B_0=300~$G case by using the 2D measured plasma density, the calculated $\mathbf{B_0}$ and for a wave travelling at $\theta=79.6^{\circ}$ with respect to the local $\mathbf{B_0}$. This value of $\theta$ is the axial average value taken from \figref[b]{VphRes}. This approximation of the 2D refractive index does not include the effects of the boundary conditions but can still provide some valuable qualitative insights, especially on-axis. \Figref{2DNref} shows the resulting 2D contours of $n$, and as expected, significant gradients of $n$ do exist. On-axis, the refractive index evolves from a plateau at $n \simeq 140$ in the centre of the column to rapid changes when $n > 150$, i.e. at $z \lesssim 18~$cm and $z \gtrsim 38~$cm. These locations roughly coincide with the WKB approximation losing its validity, i.e. where the condition $\left|\frac{\partial n / \partial z}{n}\right|^{-1} \gg \lambda_{\parallel}$ is not satisfied. This means that the wave undergo resonance at these approximate locations where a fraction of the wave is reflected and another is absorbed by the resonance \cite{stix1992waves}. While it is not possible to estimate the ratio of reflected versus absorbed wave due to the limited data available, this might however explain the wave axial damping which is especially pronounced for $z > 50~$cm \cite{takahashi2016standing}.

Interestingly, \figref{2DNref} features a conical front of resonance ($n \rightarrow \infty$) beyond which $n$ is purely imaginary, the wave is evanescent and could be subject to cut-offs ($n \rightarrow 0$). Regions where $n$ is imaginary roughly coincide with loci in \figref[b]{BzAmp2D} where $|B_0|$ is at the noise level. The fact that the conical front is non-continuous is an artefact due to the discrete measurement steps. The measured phase shifts having the characteristics of travelling waves, wave reflections at the loci of resonance are likely weak and wave absorption possibly dominates between the resonance and the cut-off regions where signs of heated electrons will be looked for in a future work.

\begin{figure}[!h]	
\begin{center}
\includegraphics[width=7.0cm]{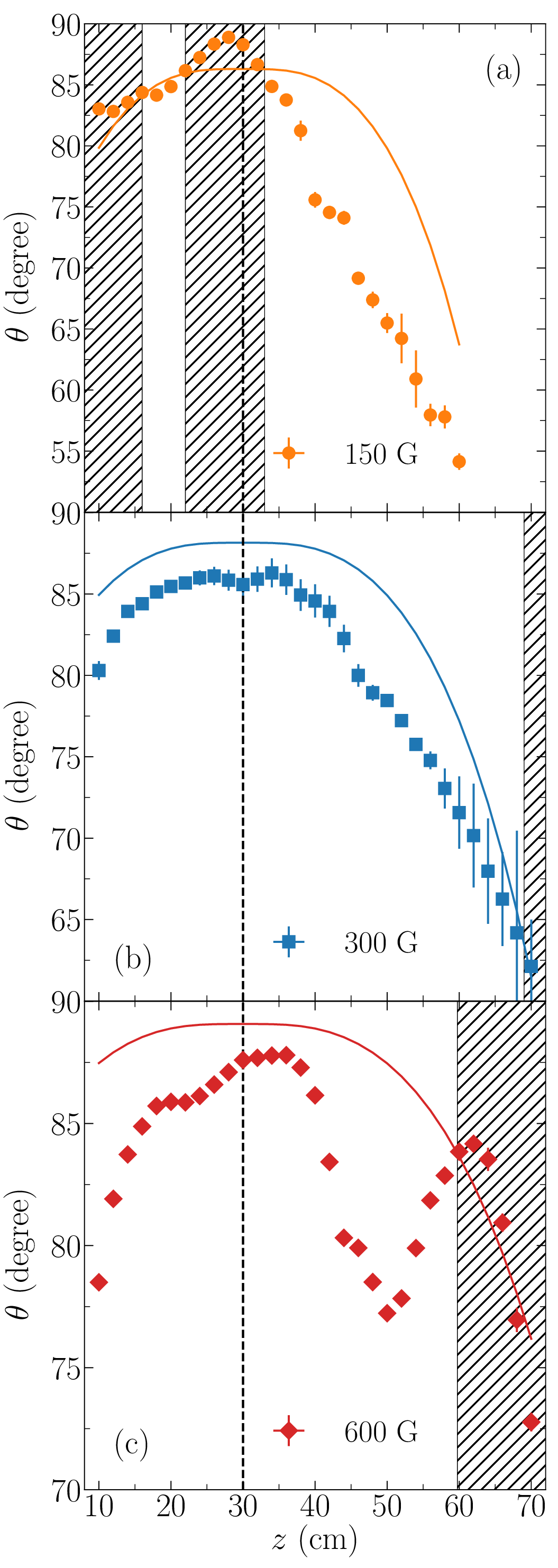}
\end{center}
\caption{Axial variation of the wavenumber angle $\theta$ (markers) and its resonance angle $\theta_{\rm res}$ (continuous line) for $B_0$ increasing from 150$~$G to 600$~$G (a)-(c). The shaded areas mark the regions where $\theta \geq \theta_{\rm res}$.}
\label{fig:VphRes}
\end{figure}

Next, the values of $\theta$ on-axis calculated from the local values $k_{\perp}$ and $k_{\parallel}$ are compared with the resonance angle $\theta_{\rm res}$ in \figref{VphRes}. Uncertainties on $\theta$ are obtained from the ones on $k_{\perp}$ and $k_{\parallel}$. At several loci, $\theta \geq \theta_{\rm res}$, most notably $z\simeq 30~$cm for $B_0=150~$G and for $z > 60~$cm when $B_0=300~$G and $600~$G. The wave locally approaching the $v_{\phi}$ resonance angle can therefore explain the anomalous ``U'' shaped profile of $|B_{\parallel}|$ when $B_0=150~$G (see \figref{BzAmp2D} and \figref[a]{WaveFeatIncB0}) as well as contribute to the wave damping for $z > 60~$cm in the other two $B_0$ cases. $\theta \geq \theta_{\rm res}$ could be the reason for $n \rightarrow \infty$ in \figref{2DNref}, suggesting that the measured wave features are consistent with the properties of the plasma medium. 

\subsection{Energy conservation}

Reflections and resonances might account for some of the wave damping, especially downstream of the magnetic mirror throat, yet, the damping of $|B_{\parallel}|$ for $z<30~$cm is still unexplained when $B_0 > 150~$G. One particularity of the discharge has been previously overlooked: its converging-diverging geometry. From \figref{BzAmp2D}, the magnetically confined plasma appears to act on the wave as a symmetric funnel of varying cross-sectional area. As such, conservation of the wave's intensity should be considered as it propagates downstream from the antenna. The intensity is equal to the time-average of the Poynting vector normal to the cross-section which is equal to the group velocity $v_{\rm g}$ times the energy density (which is $\propto B^2$) \cite{stix1992waves}. In an uniform medium, $v_{\rm g}$ would be constant and with a decreasing cross-sectional area one would expect the energy density to locally increase. This is not what is observed.

\begin{figure}[!h]
\begin{center}
\includegraphics[width=8.25cm]{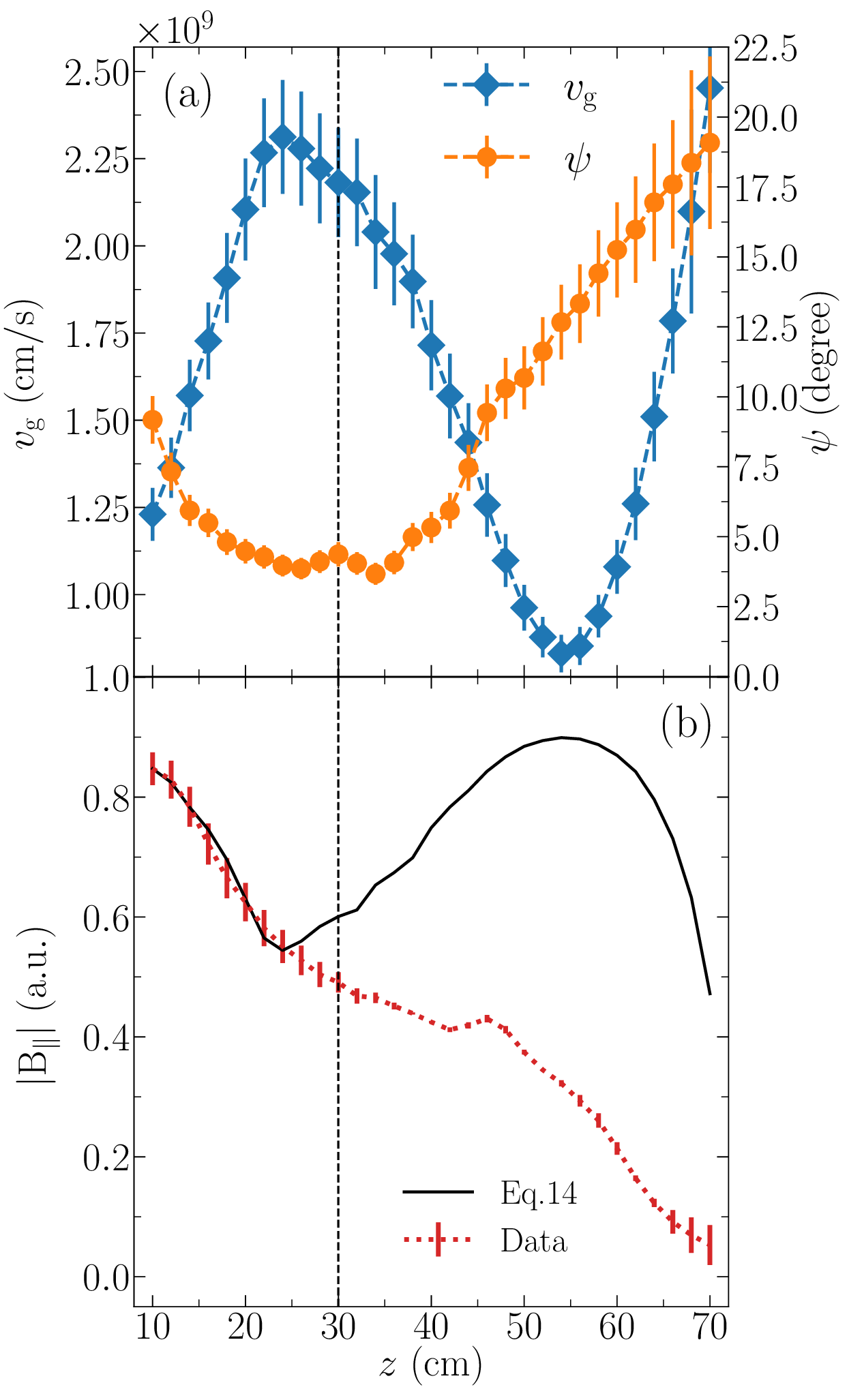}
\end{center}
\caption{Magnitude of the group velocity vector $v_{\rm g}$, its angle $\psi$ with respect to $\mathbf{\hat{z}}$ for $B_0=300~$G (a). Comparison between the measured amplitudes $|B_{\parallel}|$ (dotted line) and the amplitude $|B_{\parallel}|$ from energy conservation alone (continuous line) for $B_0=300~$G (b).}
\label{fig:ConsEnergyVg}
\end{figure}

In the present case, the medium is non-uniform and $v_{\rm g}$ varies. From \Eqref{HeliconWK}, the components of $v_{\rm g}$ in polar coordinates $(\mathbf{\hat{k}},\mathbf{\hat{\theta}})$ are
\begin{equation}
\cases{v_{\rm g k} = 2 k c^2 \omega_{\rm ce} \cos{\theta} /  \omega_{\rm pe}^2 ~,\\
v_{\rm g \theta} = - k c^2 \omega_{\rm ce} \sin{\theta} /  \omega_{\rm pe}^2 ~. \\}
\end{equation}
\Figref[a]{ConsEnergyVg} shows the magnitude of $v_{\rm g}$ and $\psi$ on-axis for $B_0=300~$G, with $\psi$ being the angle between $\mathbf{v_{\rm g}}$ and $\mathbf{B_0}$. The group velocity vector stays nearly parallel to $\mathbf{\hat{z}}$ and its magnitude is symmetrical with respect to the solenoids. The direction of the Poynting vector is approximately along $\mathbf{\hat{z}}$, and for the intensity to be conserved it is expected that as $v_{\rm g}$ increases towards $z=30~$cm, the wave energy density proportional to $B^2$ should decrease. This situation is analogous to a Venturi effect where the energy density is a proxy for the pressure of a fluid flowing at $v_{\rm g}$ through a constriction. As such, it is tempting to fit a Bernoulli-like equation of the form
\begin{equation}\label{eq:venturi}
	B(z_2)^2 - B(z_1)^2 = C \left( v_{\rm g}(z_2)^2 - v_{\rm g}(z_1)^2 \right) ~,
\end{equation} 
to the axial profile of $|B_{\parallel}|$. Here $z_1$ and $z_2$ are two axial positions and $C$ is an arbitrary constant. It is noted that for a $m=0$ mode, $B_{\perp}=0$ on-axis, and it is reasonable to expect that $B \simeq B_{\parallel}$. Taking $z_1 = 10~$cm, \Eqref{venturi} was fitted to the measured profile in \figref[b]{ConsEnergyVg}. It can be seen that this equation reproduces the wave damping up to $z\simeq 25~$cm beyond which the amplitude would increase again, as per the fluid analogue. This suggests that the local increase in the wave amplitude around $z=50~$cm for $B_0 > 150~$G could be a result of a Venturi-like effect combined with local wave resonance and reflection.

\section{Conclusions}

In this study, the 2D features of rf magnetic waves excited by a single loop antenna are characterised and identified as an $m=0$ helicon mode throughout the funnel-shaped magnetised plasma. Two-dimensional mappings of the wave amplitude show that the wave is somewhat guided by the converging-diverging plasma column. On-axis, excellent agreement is found between the measured wavelengths and the $m=0$ helicon dispersion relation including the plasma column radial boundary condition. This is verified at the local and global levels of description of the wave, for all reported values of $B_0$, and is an interesting result given the use of such a simple model on a highly non-uniform plasma. Analysis to find potential wave-plasma couplings to account for the observed wave damping on-axis revealed that neither previously described collisional nor collisionless mechanisms seem to play a significant role. The potential role of instability-driven anomalous diffusions in increasing the effective collision frequency should be further explored.

An interesting result in accordance with wave theory, is that some level of damping could be explained by local resonances of the wave at loci where its wavevector takes an angle greater than the phase velocity resonance cone. There are good correlations for the $B_0=150~$G case explaining the ``U'' shape wave axial amplitude profile, and for $B_0\neq 150~$G cases to explain the severe damping at both extremities of the plasma column. 

An unexpected outcome is that the wave damping from $z=10~$cm to the magnetic funnel can be modelled by a Bernoulli-like equation. As the cross-sectional area of the funnel decreases, the group velocity is seen to increase, resulting in a decreasing wave energy density. This does not constitute a novel helicon wave property, but rather is a feature emerging from the specific discharge geometry and the inhomogeneous plasma density. This raises the question of whether considerations of wave energy conservation and wave resonances have been sometimes overlooked in past studies which considered helicon wave damping. For the conditions reported here, phase velocity resonance is the only identified mechanism through which helicon waves could be a vector of energy transfer between the antenna and remote electrons. However, the precise coupling of such resonances are still unclear, as is the question of whether the locally deposited power is significant to the global plasma dynamics. 

Further measurement campaigns to record $B_{\perp}$ and numerical modelling efforts are needed to quantitatively assess the different wave-plasma interactions first explored in this study. Nevertheless, the data suggest that helicon waves might only play a marginal role in the remote plasma generation in the present experimental conditions. It is possible that magnetically enhanced capacitive / inductive, as well as edge waves heating processes, play determinant roles in the global power and particle balances and should be further investigated. This is also supported by the observed invariability in the plasma characteristics when changing the antenna from a double-saddle to a single-loop one, since these antennas are known to best couple to different helicon azimuthal modes ($m=1$ and $m=0$, respectively) \cite{filleul2021characterization,chabert2011physics}.

Finally, this work provides an example of a situation where high magnetic field intensities and moderate plasma densities ($10^{11} - 10^{12}~\rm cm^{-3}$) result in unambiguously characterised helicon waves in blue-mode discharges whilst contributions of the waves to the plasma generation could not be identified. Since waves are often assumed to play a role in the power balance of so-called helicon devices, these results make a case for the importance of rigorous analyses of the wave features and coupling to the plasma. The lack of such analysis can induce confusion in distinguishing between plasma generation owing to magnetically enhanced capacitive / inductive mechanisms and wave-heated regimes.

\section*{Data availability statement}

The data that support the findings of this study are available from the corresponding author upon reasonable request.

\section*{Supplementary Material}

An animation of \figref{2DTravellingWave} is available as supplementary material.

\section*{Acknowledgments}

The authors would like to thank Jim Chung for helping with the B-dot measurements during their visit to Aotearoa, as well as Philippe Guittienne for the useful conversations. This work was partially supported by the New Zealand
Space Agency under Grant No. MBIE$\#$00008060.

\section*{Conflict of Interest}

The authors have no conflicts to disclose.

\section*{References}

\bibliographystyle{unsrt}
\bibliography{./References.bib}

\end{document}